\documentclass[%
preprint,
amsmath,amssymb,
aps,
pre,
]{revtex4-2}

\usepackage[colorlinks,bookmarksopen,bookmarksnumbered,citecolor=red,urlcolor=red]{hyperref}

\usepackage{graphicx}
\usepackage{natbib}
\usepackage{color}
\usepackage{latexsym}
\usepackage{subfig}
\usepackage{natbib}
\usepackage{upgreek}
\usepackage{mathrsfs}
\usepackage{float}
\usepackage{caption}
\usepackage{soul}
\usepackage{dsfont}
\usepackage{textcomp}
\usepackage{stmaryrd}
\usepackage{amsmath}
\usepackage{lipsum}                     
\usepackage{xargs}                      
\usepackage[pdftex,dvipsnames]{xcolor}  
\usepackage[colorinlistoftodos,prependcaption,textsize=small]{todonotes}

\newcommandx{\eyal}[2][1=]{\todo[linecolor=red,backgroundcolor=red!25,bordercolor=red,#1]{#2}}

\newcommandx{\anirban}[2][1=]{\todo[linecolor=blue,backgroundcolor=blue!25,bordercolor=blue,#1]{#2}}

\def\be{\begin{equation}}
\def\ee{\end{equation}}
\def\ee{{\rm e}}

\usepackage{cleveref}

\crefformat{section}{Sec. #2#1#3} 
\crefformat{subsection}{Sec. #2#1#3}
\crefformat{subsubsection}{Sec. #2#1#3}

\begin{document}

\preprint{APS/123-QED}

\title{A Swarm Coherence Mechanism for Jellyfish}

\author{Erik Gengel}
\email{egiu@gmx.de}
\affiliation{Department of Geophysics, Porter school of the Environment and Earth Sciences, Tel Aviv University,
Tel Aviv 69978, Israel.}

\author{Zafrir Kuplik}
\affiliation{The Steinhardt Museum of Natural History, Tel Aviv University, 12 Klausner Street, Tel Aviv}

\author{Dror Angel}
\affiliation{The Leon Recanati Institute for Maritime Studies, University of Haifa, Mount Carmel, Haifa 3498838, Israel}

\author{Eyal Heifetz}
\email{eyalh@tauex.tau.ac.il}
\affiliation{Department of Geophysics, Porter school of the Environment and Earth Sciences, Tel Aviv University, Tel Aviv 69978, Israel.}

\date{\today}

\begin{abstract} 
We present a model of active Brownian particles for the process of jellyfish swarm formation. The motivation for our analysis is the phenomenon of jellyfish \textcolor{black}{blooming} in natural habitats and \textcolor{black}{clustering of jellyfish} in laboratory tanks which follow from \textcolor{black}{an interplay of} physical and behavioral mechanisms. We reach the conclusion that \textcolor{black}{phase separation at low jellyfish density is induced by behavioral reactions to environmental conditions}. Dense regions are then maintained by self-induced stimuli which result in a recruitment process. Our results agree with the biological fact that jellyfish exhibit an extreme sensitivity to stimuli in order to achieve favorable aggregations. Based on our \textcolor{black}{model}, we are able to provide a clear terminology for future experimental analysis of jellyfish swarming and we pinpoint potential limitations of tank experiments.
\end{abstract}

\maketitle

\section{Introduction}

Scyphozoans are fascinating creatures that are among the oldest multi-cellular species on earth. They feature a complex life cycle involving a benthic polyp stage and the planktonic stage of the medusa (\textit{jellyfish}) \cite{van2014origin, helm2018evolution, lotan1992life}. Scyphozoans are known to endure environmental conditions (such as hypoxia) that are harsh for many marine species, including their predators \cite{miller2012environmental, l2020fewer}. This ability has lead to an extensive debate about possible ecological and economic threats placed by jellyfish in oceans, transformed due to climate change and human actions \cite{attrill2007climate,schrope2012attack,streftaris2006alien,nakar2011economic,pitt2009influence,angel2016local, purcell2007anthropogenic, brodeur2016ecological, galil2012truth, brodeur2008rise}. 

One of the most recognized hallmarks for an increased abundance of jellyfish are massive
jellyfish blooms in the open sea and in coastal waters during certain periods of the year \cite{edelist2020phenological}. \textcolor{black}{These blooms are typically characterized by a higher jellyfish density. However, since sparsely distributed jellyfish cannot sense each other and are unlikely to interact, a basic question is how blooms emerge in the first place.}

\textcolor{black}{In fact, the transition from a homogeneous distribution to some kind of (inhomogeneous) order has been studied extensively for a variety of organisms and is, in more general terms, known as phase separation of active matter \cite{elgeti2015physics, cavagna2013diffusion, calovi2014swarming, ramaswamystatactive2010,toner1995long, adorjani2024motility, gonnella2015motility}. Often the emergent patterns are a result of already high density and a relatively pronounced motility of single agents. Similarly, emerged blooms are the final stages of phase separation in a jellyfish population. However, since jellyfish usually are sparsely distributed, processes at high agent densities (steric interactions, alignment, jamming) cannot be the cause of the transient towards blooming. Our main hypothesis is that behavioral reactions of jellyfish are the drivers of bloom formation. Instead, passive mechanisms like motility induced phase separation or jellyfish induced changes of the hydrodynamic properties in the large-scale currents are still negligible.} 

\textcolor{black}{Most of the existing jellyfish swarm models do not address the details of these initial transients despite being the cause for the later problems with blooms.} Instead, swarms are represented either by phenomenological models  \cite{brown2002forecasting,ruiz2012model, decker2007predicting,ramondenc2020probabilistic} or by passive tracers \cite{edelist2022tracking, prieto2015portuguese,el2020modelling, nordstrom2019tracking}. \textcolor{black}{Clearly, such simplifications have their merit as jellyfish often appear in large numbers and are almost entirely passive swimmers on the scale of ocean currents. However, the propulsion of single jellyfish is well understood \cite{park2014simulation, dular2009numerical,hoover2021neuromechanical,gemmell2013passive, costello1995flow,pallasdies2019single, dabiri2005flow}. Those results suggest that across diverse strategies of swimming, jellyfish are the most efficient swimmers in the oceans \cite{costello2021hydrodynamics}. Additionally, observations have shown that active swimming of jellyfish has an effect on their distribution in coastal waters and tanks \cite{fossette2015current, malul2019levantine}. A few models have incorporated a rudimentary type of active behavior or more than one environmental driver of transport into a tracer model to make predictions \cite{north2008vertical, macias2021model, malul2024directional}. However, those models are mostly explanatory for the observed data at hand or remain limited in scope. A further need for more generic modelling is given by the notorious problems with behavioral changes of jellyfish in tanks due to confinement. In those cases, it would be beneficial to have a model at hand that allows to differentiate genuine and artificial behavior.}

What are the ingredients for such a model? In the open sea, many jellyfish species are known to migrate vertically on a diurnal basis such that large amounts of jellyfish can be situated in a relatively narrow layer of the surface water for hours. There, a swarm gets mainly affected by surface currents such that a two-dimensional \textcolor{black}{model} of jellyfish swarming can already provide insight into the distribution of jellyfish in those periods of the day \cite{gershwin2014dangerous, petersen2019experimental, cressman2004eulerian}. On the contrary, a swarm usually dives \textcolor{black}{into deeper water to avoid waves during storms}. In those situations, the three-dimensional large-scale flow is needed to understand jellyfish transport \textcolor{black}{on shorter time scales. On long time scales it again appears convenient to neglect vertical migration as it represents a fast fluctuation, leaving the horizontal transport as a main component of jellyfish dispersion \cite{sundermeyer1998lateral}.} In bays, estuaries and in experiments, jellyfish have been observed to react to a variety of different environmental conditions: They swim actively against currents \cite{fossette2015current, malul2019levantine} or they follow the direction of the surface waves \cite{malul2024directional}. Moreover, distributions of jellyfish are affected by other environmental drivers, such as salinity and temperature \cite{zhang2012associations, heim2019salinity, dror2023rising}, solar insulation \cite{bozman2017jellyfish, hamner1994sun} and food availability \cite{arai1991attraction}. 
\textcolor{black}{These drivers can be thought of as clues to orient during bloom formation, when the single medusae are still sparsely distributed.} 

In particular preying and sexual reproduction \cite{hamner2009review} can be hypothesized to have a profound effect on jellyfish transport beyond environmental drivers: Observations suggest that jellyfish are attracted to chemicals that are secreted by their planktonic prey \cite{arai1991attraction, hays2012high}. In the open sea, the distribution of this prey gets elongated along fronts of mesoscale eddies \cite{lehahn2007stirring, verma2021lagrangian}. Thus, already a persistent active swimming effort on a relatively small scale, guided by a chemical gradient causes jellyfish to cross stream lines, causing their surrounding flow to changes. Similarly, there exists some limited evidence that jellyfish react to sexual pheromones \cite{garm2015mating}. Such a behavior would indeed compare to that of other cnidarians \cite{tarrant2005endocrine}. While such a bio-chemical stimulus remains to be understood in more detail, it is a fact that during periods of sexual reproduction, jellyfish use dense swarms to their advantage, to increase the success of fertilization. It is therefore straightforward to hypothesize that pheromones act in combination with pressure fluctuations and released oocytes as a \textit{self-induced} recruiting stimulus. \textcolor{black}{This stimulus gains behavioral importance when the faint effects of single medusae on the environment add up to a macroscopic stimulus at a higher local jellyfish density.}

\textcolor{black}{Our model combines some of the aforementioned behavioral mechanisms} together with first principles of computational neuroscience and physics into a comprehensive framework in order to understand active jellyfish swimming \cite{gengel2023physics}. It is rooted in the theory of active matter and represents jellyfish as active Brownian particles (ABPs) \cite{romanczuk2012active, schweitzer2003brownian}. In our case, these ABPs are equipped with a minimal set of jellyfish-like properties, in spirit of the general idea of active-matter theory, to describe the emergence of collective phenomena in an ensemble of actively moving agents based on the interaction and decision rules. 

We discuss the self-induced recruitment of sparse ABP-jellyfish and their swarm formation in a confined 2D flow. This scenario relates to the mostly observed quasi-surface swimming of jellyfish in the oceans and to the setup of jellyfish swimming in a tank \textcolor{black}{ \cite{hansson1995behavioural,mackie1981swimming, hansson1997capture,titelman2006feeding,bailey1983laboratory, albert2011s}}. Our paper is organized as follows: In Sec.~\ref{sec: jellyfish model} we introduce the model and its reasoning. In Sec.~\ref{Sec: main methods} we describe simulation and analysis methods. \textcolor{black}{Results and discussion are presented in Sec.~\ref{Sec: results}}, followed by concluding remarks in Sec.~\ref{sec: conclusions}.

\section{Jellyfish model}\label{sec: jellyfish model}

\begin{figure}
    \centering
    \includegraphics[width=\columnwidth]{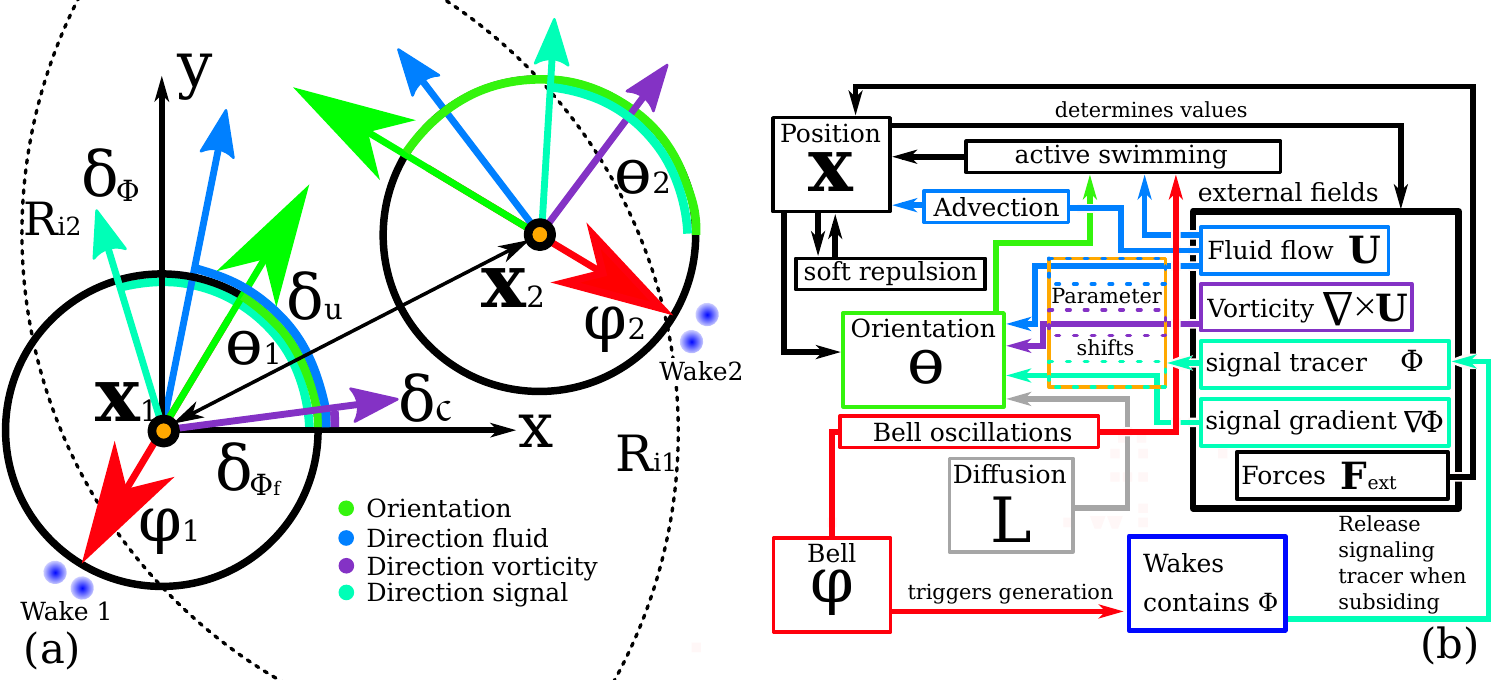}
    \caption{Depicted is a schematic of the ABP-jellyfish dynamics. (a): Agent-agent interaction and environmental stimuli with colors corresponding to panel (b). The interaction radii $R_{i;1,2}$ are depicted by dashed lines. (b): Schematics of the model for jellyfish swarming. This model considers the flow $\mathbf{U}$, its vorticity ${\cal C}$ and a self-induced signaling tracer $\phi$ as inputs. Additionally, the model incorporates turning near walls and repulsion close to walls. The external inputs are subject to parameter shifts.}
    \label{fig: schematic model_2}
\end{figure}

Active jellyfish swimming is realized by the bell oscillations of a medusa which causes short-range fluid perturbations and a wake vortex to emerge. We assume that the wakes simultaneously transport a self-induced tracer which is a blend of physical, chemical and cellular stimuli \cite{dabiri2005flow, costello2021hydrodynamics}. Accordingly, jellyfish decision making involves four aspects:
\begin{itemize}
    \item[A] The internal dynamics of individuals causes the ambient water to move.
    \item[B] The flow field in close proximity and the wake field of an individual act as transfer media of information and lead to volume exclusion.
    \item[C] The reactions and the decisions of jellyfish emanate based upon the received information.
    \item[D] Changes of the environmental conditions can be induced by the presence of jellyfish.
\end{itemize}
Accordingly, we propose to model a jellyfish $j$ as an ABP with a horizontal position $\mathbf{x}_j=[x,y]_j$, an orientation $\theta_j$ and an internal state that accounts for the nonlinear oscillation of the bell. \textcolor{black}{We consider only an effective signaling tracer $\phi$ that is induced by the jellyfish and which gets advected with the flow. The resulting chemical communication is well known for bacteria \cite{miller2001quorum} and ant colonies \cite{couzin2003self, moussaid2009collective}. With application in mind we avoid any further hydrodynamic couplings of the environment to the presence of jellyfish. On the one hand, because, swimming of mature jellyfish takes place at Reynolds numbers of $10^4$ which is roughly four orders of magnitude smaller than the Reynolds number associated with kilometer scale eddies. On the other hand, jellyfish are almost neutrally buoyant \cite{yang2018rowing, suzuki2018mechanisms} and pressure anomalies around their bell and in the wake are of the order of a few Pascal only \cite{gemmell2013passive}. Thus, prior to blooming, jellyfish induced fluctuations of neither density nor viscosity are significant in the bulk flow.} Figure~\ref{fig: schematic model_2} depicts the \textcolor{black}{resulting} jellyfish model in full detail. \textcolor{black}{In panel (a) the microscopic directional inputs on the level of single agents are shown. Panel (b) shows the flow diagram of the model mechanism discussed next.}

\subsubsection{Bell Oscillations}

Jellyfish possess a simplistic yet highly efficient neuronal net that is distributed over the bell \cite{albajes2011jellyfish, satterlie2011jellyfish,garm2006rhopalia,mackie2004central, pallasdies2019single, berner2021synchronization}. \textcolor{black}{Most importantly, the oscillation of the bell can be associated with a stable limit cycle of the network dynamics $\dot{\mathbf{q}} = \mathbf{F}(\mathbf{q})$. Along the cycle $\mathbf{q}_0(\varphi)$, the dynamic is described by a $2\pi$-periodic phase $\varphi(\mathbf{q})$ with dynamics $\dot{\varphi} = \omega$. An additional perturbation $\mathbf{P}(t)$ results in 
\begin{equation}
\dot{\mathbf{q}} = \mathbf{F}(\mathbf{q})+\mathbf{P}(t) \qquad \Rightarrow \qquad \dot{\varphi} = \nabla_{\mathbf{q}} \varphi(\mathbf{q}) \left[ \mathbf{F}(\mathbf{q}) + \mathbf{P}(t)\right]  = \omega + Q(\varphi,t)\; . \label{eq: phase unperturbed}
\end{equation}}
Such a \textcolor{black}{phase reduction} keeps essential features of the bell oscillator and the essence of most fluid-dynamics studies on jellyfish, it reduces greatly the computational complexity and it allows to build a data-driven model of the bell dynamics \cite{Kralemann_etal-13, pikovsky2001synchronization,nakao2021phase, schwabedal2012optimal}. 

\textcolor{black}{In Eq.~\eqref{eq: phase unperturbed} the unperturbed bell dynamics ($\mathbf{P} \equiv 0$) is represented by a constant average frequency $\omega_j=\nabla_{\mathbf{q}} \varphi(\mathbf{q}_0) \mathbf{F}(\mathbf{q_0})$} that can vary according to the size of individuals, species and seasons. Here, we assume a constant bell frequency of $\omega_j=\Omega$. The phase maps to different events in a pulsation cycle. We adapt the convention that a closed bell is associated with a phase of $\varphi=0$. \textcolor{black}{The coupling function $Q(\varphi,t)$ represents the effect of perturbations $\mathbf{P}$. In case of weak perturbations it is given by $Q(\varphi,t)=\mathbf{Z}(\varphi)\cdot \mathbf{P}(t)$ where $\mathbf{Z}=\nabla_{\mathbf{q}} \varphi(\mathbf{q}_0(\varphi))$ is the $2\pi$-periodic phase response curve that contains all characteristics about the bell response to an external perturbation \cite{schultheiss2011phase}. For stronger perturbations, a more generic form of system reduction is needed \cite{shirasaka2017phase, mauroy2018global}. We assume that $\mathbf{P}$ during bloom formation is a random input resulting from turbulence such that long term averages of $Q(.)$ either vanish or can be absorbed into the constant bell frequency. Thus, we let $Q(.) \equiv 0$.}

\subsubsection{Wake Dynamics}

Jellyfish eject wakes that consist of starting and stopping vortex rings and vortex super structures \cite{sahin2009numerical, dabiri2005flow, herschlag2011reynolds, gemmell2015control}. The main thrust is produced when the bell contracts and closes, generating the starting vortex. In our model this corresponds to a bell phase of $\varphi= 0$. \textcolor{black}{Over the succession of many oscillation cycles, the generated vortex street subsides into the surrounding medium and forms a continuum of bell turbulence. We assume that this initial transient is fast such that the incentive $\phi$ gets released into the bulk flow immediately. Accordingly, $\phi$ accounts not just for pheromone release during sexual reproduction but also for turbulence effects which have a behavioral effect but are too weak to influence the bulk flow.}

\subsubsection{Positional Dynamics}

\begin{figure}
    \centering
    \includegraphics[width=\columnwidth]{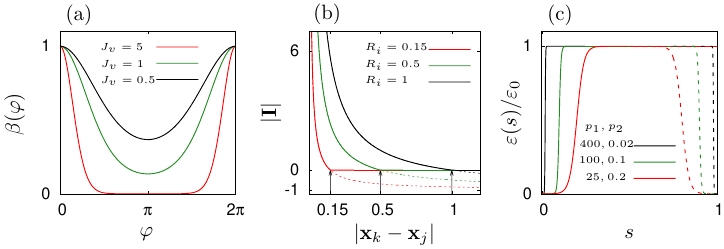}
    \caption{Panel of coupling functions. (a): Phase dependence of the velocity magnitude Eq.~\eqref{eq vel magnitude} for parameters $J_v\in[5,1,0.5]$; (b): Radial dependence of the volume exclusion term in Eq.~\eqref{eq: repulsion field} for different interaction radii $R_i\in[0.15,0.5,1]$. Dashed lines indicate the omitted attractive part of the interaction; (c): Normalized coupling parameters $\varepsilon(s)/{\varepsilon_0}$ Eq.~(\ref{eq: UC ignorance}, \ref{eq: epsphi2 response}) for three sets of parameters $p_1$, $p_2$. \textcolor{black}{Solid lines indicate responses associated with velocity and turbulence in Eq.~\eqref{eq: UC ignorance} ($s=|\mathbf{U}|$ or $s=|{\cal C}|$). The response with the signaling tracer ($s=\phi$) Eq.~\ref{eq: epsphi2 response} is similar at small stimuli and deviates for $s>0.5$, indicated by dashed lines.}}
    \label{fig: couplings}
\end{figure}

\textcolor{black}{Jellyfish movement is described by an equation of the form 
\begin{equation}
    m \Ddot{\mathbf{x}}_j = \gamma_1 \left( \mathbf{U}_j - \dot{\mathbf{x}}) \right) + \gamma_2 \mathbf{V}_j + \mathbf{F}, \qquad \mathbf{V}_j=V(\varphi_j) \hat{e}_j(\theta_j), \qquad \mathbf{F} = \frac{{\cal F}_0}{N^a_j}\sum_{k\neq j}^{N^a_j} \mathbf{I}(\mathbf{x}_k, \mathbf{x}_j) + \mathbf{F}_{\text{ext}}\; .
\end{equation}
Here, $m$ is the mass of jellyfish, ${\cal F}_0$ is the force scaling of interaction and $\gamma_1 =\gamma_2 = \gamma$ are friction coefficients such that ${\cal F}_0/\gamma = 1$ and $\mathbf{f}_{\text{ext}} = \mathbf{F}_{\text{ext}}/\gamma$. For our modelling, we assume that $m \Ddot{\mathbf{x}}$ is negligible such that} the positional dynamics \textcolor{black}{is}
\begin{equation}
\dot{\mathbf{x}}_j = \mathbf{U}_j + V(\varphi_j) \hat{e}_j(\theta_j) + \frac{1}{N^a_j}\sum_{k\neq j}^{N^a_j} \mathbf{I}(\mathbf{x}_k, \mathbf{x}_j) + \mathbf{f}_{\text{ext}}\; , \label{eq: positional dynamics}
\end{equation} 
Eq.~\eqref{eq: positional dynamics} takes into account advection according to the local background flow $\mathbf{U}_j=\mathbf{U}(\mathbf{x}_j)$, volume exclusion and active swimming into direction $\theta_j=\text{arg}(\hat{e}_j)$ with a speed 
\begin{equation}
 V(\varphi_j) = \left(V_0 + \frac{V_1 |\mathbf{U}|}{V_2 + |\mathbf{U}|} \right)\beta(\varphi_j), \qquad \beta(\varphi_j) = e^{J_v(\cos(\varphi_j)-1)} \; . \label{eq vel magnitude}
\end{equation}
The function $\beta(.)$ mimics the pulsatile motion of jellyfish (see Fig.~\ref{fig: couplings} panel (a)). Its shape depends on the parameter $J_v>0$. For large $J_v$, a jellyfish comes to a full stop in its bell cycle while for small $J_v$, an agent maintains a certain level of speed \cite{park2014simulation, dular2009numerical}. \textcolor{black}{The baseline swimming speed is $V_0$.} The additional nonlinear dependence of the velocity magnitude on the magnitude of the mean flow, $|\mathbf{U}|$ takes into account that jellyfish are able to swim faster against the direction of flow \cite{fossette2015current, malul2019levantine}. The response is quasi-linear with slope $V_1/V_2$ at small flow velocities and it saturates at large flow velocities to $V_1$ \cite{gengel2023physics}. \textcolor{black}{Accordingly, $V_0$ and $V_1$ determine the equilibrium flow speed at which a single individual jellyfish stands still when swimming against th flow.} Parameters of this mechanism are listed in Tab.~\ref{tab: params inputs2} \textcolor{black}{and have been adapted to the experimental results for \textit{Rhopilema nomadica} reported in \cite{malul2019levantine}. There, the jellyfish swim with a speed of $\langle V \rangle_t=0.067$\ m\ s$^{-1}$ in a quiescent fluid while at a background flow speed of $|\mathbf{U}| = 0.045$\ m\ s$^{-1}$, \textit{R.~nomadica} achieves an average speed of $\langle V \rangle_t = 0.083$\ m\ s$^{-1}$ against the flow. From this information, we can estimate $V_0$, $V_1/V_2$ and $V_1$. $\langle.\rangle_t$ indicates averaging over time.}

$N^a_j$ is the number of ABP-jellyfish found within a distance of $R_i$ around the jellyfish $j$. We model the interaction at short distances by a static and rotational symmetric repulsion 
\begin{equation}
\mathbf{I}(\mathbf{x}_j,\mathbf{x}_k) = \frac{\mathbf{x}_k-\mathbf{x}_j}{|\mathbf{x}_k-\mathbf{x}_j|} - R_i \frac{\mathbf{x}_k-\mathbf{x}_j}{|\mathbf{x}_k-\mathbf{x}_j|^2} \; , \label{eq: repulsion field}
\end{equation}
acting in between individuals $j$ and $k$ at distances $|\mathbf{x}_k-\mathbf{x}_j|< R_i$ \cite{aranson2013active,mognetti2013livin, weeks1971d} (see Fig.~\ref{fig: couplings} panel (b)). At the edges of the computational domain, the walls repel agents by a reset force $\mathbf{F}_{\text{ext}}$. 

\textcolor{black}{By assuming an overdamped dynamics Eq.~\eqref{eq: positional dynamics} and a symmetric interaction Eq.~\eqref{eq: repulsion field} we largely neglect short range turbulence, its inhomogeneities and inertia in the positional dynamics. The only inertial effect still present is given by $\beta(.)$ (see black arrows and red arrows in Fig.~\ref{fig: schematic model_2} panel (b)).}

\subsubsection{Orientation Dynamics}

\textcolor{black}{In the transient towards blooming} jellyfish react only to environmental drivers with a behavioral response. For this, it suffices to consider an externally forced angular dynamics 
\begin{equation}
\begin{aligned}
\dot{\theta}_j & = L_j(t) + E(\mathbf{U}_j,{\cal C}_j,\theta_j) - \varepsilon_{\phi}(\phi) \sin(\theta_j-\delta_{\phi}) + \pi \sin(\theta_j-\delta_{n,j})) {\cal B}(\mathbf{x}_j)
\end{aligned}\; . \label{eq: angular coupling}
\end{equation}
The angular diffusion due to the fluctuation $L_j$ mirrors the effect of self-induced turbulence on the swimming direction of jellyfish (see grey box in Fig.~\ref{fig: schematic model_2} panel (b)). What stands behind this is a neuro-mechanical feedback \cite{gemmell2015control,hoover2021neuromechanical} in which the turbulent fluid motion around a jellyfish causes a macroscopic tumbling. We assume this effect to be uncorrelated with other environmental drivers but correlated in time. Accordingly, we model $L_j$ by an Ornstein-Uhlenbeck process \cite{uhlenbeck1930theory} following the dynamics
\begin{equation}
\dot{L}_j = -\lambda_{\theta} L_j + \eta(t),\qquad \langle \eta \rangle_t = 0, \; \; \langle \eta(t) \eta(t') \rangle_t = 2D_{\theta} \delta(t-t') \; .
\end{equation}
Here, $D_{\theta}$ is the angular diffusion coefficient and $\lambda_{\theta}$ is the noise damping  \cite{polin2009chlamydomonas, fier2018langevin, saragosti2012modeling}.

The behavioral effects of the flow field $\mathbf{U}_j$ and the vorticity field ${\cal C}_j = \nabla \times \mathbf{U}_j$ (see blue and violet boxes in Fig.~\ref{fig: schematic model_2} panel (b)) are taken into account by a phase coupling function, $E(.)$
\begin{equation}
    E(\mathbf{U}, {\cal C}, \theta_j) = \varepsilon_{\mathbf{U}}(\mathbf{U}) \sin(\theta_j-\delta_{\mathbf{U}}) + \varepsilon_{{\cal C}}(|{\cal C}|) \sin(\theta_j-\delta_{{\cal C}}) \; .
\end{equation}
Herein, $\delta_{\mathbf{U}}=\text{arg}([U_x,U_y])$ and $\delta_{\cal C} = \text{arg}([\partial_x |{\cal C}|,\partial_y |{\cal C}|])$ are the directions of flow and vorticity gradient. Pattern formation due to flow and vorticity has been previously studied in combination with a prey tracer \cite{gengel2023physics} and can be extended to \textcolor{black}{environmental drivers of salinity, temperature or solar insulation} \cite{ambler2002zooplankton, hamner2009review}. 

However, once the fluid is quiescent, its directional clues should be ignored to save energy ($E(.) \equiv 0$). This situation can be easily incorporated into the angular dynamics by a parameter response of the form 
\begin{equation}
\begin{aligned}
    \varepsilon_{\mathbf{U}}(|\mathbf{U}|) &= \varepsilon_{\mathbf{U},0} {\cal T}(S_{\mathbf{U},1}, S_{\mathbf{U},2},|\mathbf{U}|), \\
    \varepsilon_{{\cal C}}(|{\cal C}|) &= \varepsilon_{{\cal C},0} {\cal T}(S_{{\cal C},1}, S_{{\cal C},2},|{\cal C}|), \\
    {\cal T}(p_1,p_2,s) &= \frac{1}{2} \Big (1+\tanh(p_1(s-p_2)) \Big) \; .  
\end{aligned} \label{eq: UC ignorance}
\end{equation}
The activation function ${\cal T}(.)$ is a reminiscence of neural field models \cite{karlik2011performance, sompolinsky1988chaos} for neurological responses. Here it allows for a complex decision making. $\varepsilon_{\mathbf{U},0}$ and $\varepsilon_{{\cal C},0}$ are the reorientation strengths. They are inversely proportional to their respective reorientation time scales. Parameter $p_1$ determines how sudden a reaction to a change in stimulus $s$ takes place and $p_2$ represents the sensitivity threshold for the stimulus (see \textcolor{black}{solid lines in} Fig.~\ref{fig: couplings} panel (c)). Accordingly, $S_{\mathbf{U},2}$ and $S_{{\cal C},2}$ set the intensity levels at which a response is switched on. We adapt $S_{\mathbf{U},1}$ and $S_{{\cal C},1}$ such that the responses resembles a simple on-off switch. \textcolor{black}{The reason for this is that jellyfish react already to small stimulus amplitudes. Accordingly $p_2$ need to be small which demands a large $p_1$ to still ensure a vanishing response at $|\mathbf{U}|=|{\cal C}|=0$ (see the orange box in Fig.~\ref{fig: schematic model_2} panel (b) and Fig.~\ref{fig: couplings} panel (c)). Accordingly, throughout this work we use a numerical value of $p_1=400$ (see Tab.~\ref{tab: params inputs2}). for all responses.} For more complex decision processes one may introduce a product ${\cal T}_1(s_1,\ldots){\cal T}_2(s_2,\ldots)\ldots$ of different response functions. 

\begin{figure}
    \centering
    \includegraphics[width=\columnwidth]{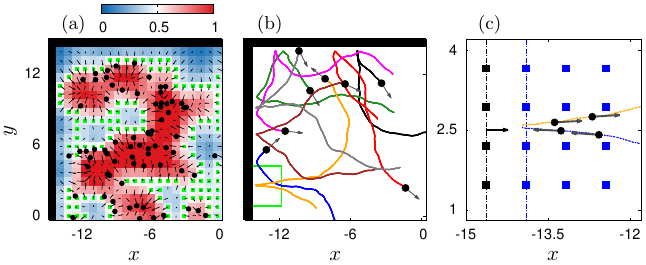}
    \caption{(a): Depicted is a snapshot of the upper left corner of the swarming domain for $D_{\phi}=0.001$\ m$^2$\ s$^{-1}$, $S_{\phi}=0.41$, $\varepsilon_{\phi,0}=0.32$\ rad\ s$^{-1}$ corresponding to Fig.~\ref{fig: patches small tank}. Jellyfish are indicated by black dots and black arrows indicate the gradient of $\phi$. Green markers indicate the region where self-induced stimulus is affecting the behavior ($\varepsilon_{\phi}>0$). (b): Shown are example trajectories of jellyfish near the left and upper walls in the absence of chemotaxis ($\varepsilon_{\phi}=0$). The green box indicates a zoom to one of the turning events in panel (c). Four example positions of the agents are shown before and after the turning event, also indicated by the blue and orange part of its trajectory. Blue squares indicate fluid cells and black squares indicate the wall elements. The horizontal arrow pointing to the right indicates the wall normal $\hat{e}_{\text{wall}}$ corresponding to $\delta_{n}=0$.}
    \label{fig: wall gradient}
\end{figure}

To model \textcolor{black}{behavioral swarm recruitment} due to the signaling tracer $\phi$ \textcolor{black}{we introduce an additional angular} coupling term of strength $\varepsilon_{\phi}$ \textcolor{black}{which} causes each agent to orient into directions $\delta_{\phi} = \text{arg}([\partial_x \phi,\partial_y \phi])$ (see Fig.~\ref{fig: wall gradient} panel (a)). In our model, the strength of this interaction changes with the ambient signaling tracer concentration according to
\begin{equation}
    \varepsilon_{\phi}(\phi) = \varepsilon_{\phi,0} {\cal T}(S_{\phi,1},S_{\phi,2},\phi){\cal T}(S_{\phi,1},S_{\phi,2}-1,-\phi) \; . \label{eq: epsphi2 response}
\end{equation}
This approach implements ignorance similar to Eq.~\eqref{eq: UC ignorance} for weak stimuli ($\phi \rightarrow 0$). At a strong stimulus ($\phi \rightarrow 1$), we also assume decoupling from $\delta_{\phi}$ because it corresponds to a situation where jellyfish are absorbed in a swarm where agent-agent interactions prevail over environmental stimuli. Upper and lower threshold are symmetrically set by $S_{\phi,2}$ \textcolor{black}{(see solid and dashed lines in Fig.~\ref{fig: couplings} panel (c)).}

Note that the dynamics Eqs.~\eqref{eq: angular coupling} depends only on the direction of the stimulus, $\delta_{\phi}$ and $\phi$ through ${\cal T}(\phi)$, in contrast to common models of chemotaxis in which $\dot{\hat{e}}(\theta)$ depends also on the magnitude $|\nabla \phi|$ \cite{gelimson2016multicellular}. The reason for this is twofold: First, jellyfish are known to react to touch where already a single capture event (for example of \textcolor{black}{a fish larva)} causes a distinct change in behavior \cite{bailey1983laboratory}. \textcolor{black}{The probability of such capture events to occur is given by $\phi$.} Second, jellyfish have a limited \textcolor{black}{physiology \cite{costello2008medusan}}. Thus, the tendency can not \textcolor{black}{have a strong effect on angular coupling.} In \cite{gengel2023physics} a scaling according to $|\nabla \phi|$ has been incorporated by means of an additional activity variable acting in Eq.~(\ref{eq vel magnitude}, \ref{eq: angular coupling}) which we avoided here to keep the essence of the \textcolor{black}{behavioral} clustering mechanisms.

Finally, jellyfish in tanks are faced with a considerable problem: They have to avoid walls to preserve their fragile bodies while their angular diffusion is much to slow. This is indicated by a persistence time $\tau_p=\lambda_{\theta}^2 / D_{\theta}=25$\ s$^{-2}$\ $0.1$\ s$^3 \sim 250$\ s \textcolor{black}{which is much larger than their ballistic time scale} $2R_i/\langle V \rangle_t= 0.3$\ m$/0.067$\ m~\ s$^{-1} \sim 5$\ s \cite{malul2019levantine, gengel2023physics, sprenger2023dynamics, gotze2010mesoscale}. Thus, jellyfish must turn actively near walls. \textcolor{black}{Similarly, jellyfish in coastal waters swim against the direction of incoming waves \cite{malul2024directional}.} A simplistic approach to include both effect is an angular coupling against the normal of walls $\hat{e}_{\text{wall}}$ with direction of $\delta_{n,j} = \text{arg}(\hat{e}_{\text{wall}})$. We consider a relatively fast rotation rate of $\pi {\cal B}(\mathbf{x}_j)$ rad~s$^{-1}$ which causes a full reflection to last for $1$\ s. This wall interaction sets in only when an agent is near a wall. We model this by a Boolean mapping $ {\cal B}(\mathbf{x}_j)$. (see Fig.~\ref{fig: wall gradient} panels (b,c) and App.~A).

It has to be emphasized that the observed values of $\tau_p$ in tank experiments depend on tank size because jellyfish are extremely sensitive to obstacles. Thus, diffusion of solitary jellyfish is best inferred from a tank significantly larger than the persistence length \textcolor{black}{$L_p = \langle V \rangle_t \tau_p = 0.067$~m~s$^{-1} 250$~s$\sim 17$\ m}. For larger medusae this requirement easily yields tank sizes of several thousand square meters. For example, in our simulations, we have used domains with edge length of roughly $2L_p$ (grid 1; 900\ m$^2$) or $5L_p$ (grid 2; $8100$\ m $^2$). In contrast, smaller tanks are suitable for inference of jellyfish swimming in confined environments like dense swarms \cite{cavagna2013diffusion,murakami2015inherent}. Consequently, it is imperative to investigate experimentally, how jellyfish interact with walls and surfaces \cite{uspal2015rheotaxis}. 

\subsubsection{Tracer Dynamics}

The signaling tracer in the fluid evolves according to the advection-diffusion equation
\begin{equation}
    \frac{D \phi}{Dt} = D_{\phi} \Delta \phi + {\cal S}\;  \label{eq: diffusion of tracers}.
\end{equation}
Here, $D/Dt \equiv \partial_t + U_x \partial_x + U_y \partial_y$ is the horizontal advective derivative and $D_{\phi}$ is the diffusion constant of $\phi$ \cite{kundufluiddyn}. 
The forcing term (see mint-green in Fig.~\ref{fig: schematic model_2} panel (b)) is  
\begin{equation}
    {\cal S}(\rho,\phi) = C_{\phi}\text{Coupling}(\rho,\textcolor{black}{\phi}) - \Gamma \phi \label{eg: sink signaling} \; .
\end{equation}
In our model the source is resembled by superposition of Gaussian bell curves tied to $N$ discrete jellyfish positions. These positions correspond to the jellyfish area density $\rho(\mathbf{x})$ at the Eulerian position $\mathbf{x}$ (see App.~B). The amplitude of the source is given by the mixing factor $C_{\phi}=\text{Volume}_{\text{bell}}/\text{Volume}_{\text{cell}}$ which is defined by the ratio of grid cell volume and bell fluid volume. $C_{\phi}$ is essentially a free parameter, determined by the vertical depth of a grid cell and the bell physiology (see Sec.~\ref{sec: params}). The coupling of the tracer to the fluid grid assumes an initial diffusive mixing due to the presence of wakes. \textcolor{black}{Tracer release is gradually reduced near $\phi=1$ as a means to ensure numerical stability (see App.~B). Indirectly, this accounts also for further potential regulation mechanisms we are not interested in here \cite{lucas2001reproduction}. Finally,} we model a simplistic exponential sink term with a decay constant $\Gamma$. The sink models the effect of sedimentation, vertical advection or microbial consumption.

\textcolor{black}{The definition of $\phi$ depends on a saturation constant $\phi_0$ which sets the value of $\phi=1$. For example, a fluid could be considered saturated with oocytes if a concentration correlates with a saturated response of jellyfish.}

\section{Methods}\label{Sec: main methods}

\subsection{Numerical Simulation}

The dynamics of ABP-jellyfish is simulated using the stochastic Heun method \cite{wilkie2004numerical}. The Diffusion of tracers Eq.~\eqref{eq: diffusion of tracers} and the fluid flow are simulated on a staggered grid with $41\times41$ grid points spanning a quadratic area of $A_{\text{tank}} = 30 \times 30$\ m (grid 1) or $A_{\text{tank}} = 90 \times 90$\ m (grid 2). These resolutions are much larger than the short-range scales of agent-agent interaction, to cover the common situation of ocean simulations which are significantly coarse grained \cite{maulik2016dynamic}. We solve the Navier-Stokes equation for a density of $\rho_f=1000$\ kg\ $m^{-3}$ and an effective viscosity of $\mu=2$\ Ns\ m$^{-2}$ on grid 2 (see Sec.~\ref{sec: counter flow}). The agent time step is set to $0.01$\ s. Data is collected every $0.5$\ s. We use a tri-linear spatio-temporal interpolation to couple the ABP-jellyfish to the Eulerian field values of $\mathbf{U}$, ${\cal C}$ and $\phi$ at time $t$ \cite{press1992numerical}. In turn, we only couple the tracer dynamics to the dynamics of agents in Eq.~\eqref{eg: sink signaling}. For this, we use spreading radii of $R_s=1$\ m (grid 1) and $R_s=2$\ m (grid 2). The software for this simulation can be found under \url{https://gitlab.com/IUFRGMP}

\subsection{Statistical Analysis}

We make use of the following statistical quantities:
\begin{equation}
\begin{aligned}
    {\cal R}_n &=& \sqrt{\langle \langle r^2 \rangle_n \rangle_N}, & \qquad \text{Local mean-square distance} & \\    
    \text{Hex}_m &=& \langle |\langle \exp{i 6\delta} \rangle_m| \rangle_N, & \qquad \text{Hexagonal order parameter} & \\
    \Phi &=& \frac{\langle \phi \rangle_N}{\langle \phi \rangle_{\text{tank}}}, & \qquad \text{Tracer ratio} \\
    {\cal N} &=& \langle K \rangle, & \qquad \text{Average cluster size}
\end{aligned} \label{eq: statistical quantities}
\end{equation}
Herein, $\langle . \rangle$ denotes ensemble averages. Lower case indices denote average over the $n$ and $m$ nearest neighbors and index $N$ indicates averaging over all agents. ${\cal R}_n$, Hex$_m$ and ${\cal N}$ are empirically accessible quantities while $\Phi$ serves here to elucidate the process of clustering \textcolor{black}{in phase B)}. For ${\cal R}_n$ and Hex$_m$ we adapted a two-stage averaging since the swarm generally consists of several, disconnected, clusters (see Fig.~\ref{fig:MSQdist tracer eps dependence } panels (b-e)). ${\cal R}_n$ and Hex$_m$ are additionally normalized by their values near $t=0$ to make visible subtle changes relative to the initial state. We average our model results over $M$ runs and locally over time, using a Gaussian kernel-density smoother with a time window of $20$\ s \cite{hastie2009kernel}. For the local ensemble averages we use the $n=6$ and $m=24$ nearest neighbors. In the following, we drop these indices for simplicity. 

\begin{figure}[!btp!]
    \centering
    \includegraphics[width=\columnwidth]{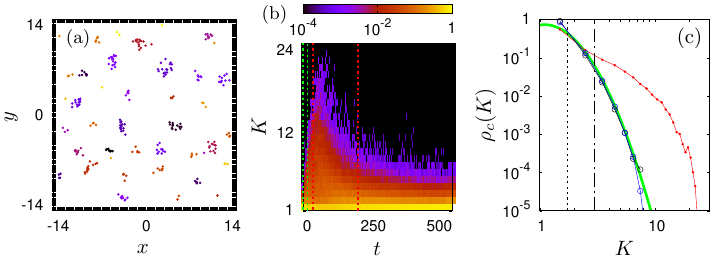}
    \caption{(a): Depicted is an exemplary decomposition of a jellyfish swarm into sub clusters for $R_c=0.5$\ m. Dots of same color belong to the same cluster. The snapshot is taken at $100$\ s from the dynamics in Fig.~\ref{fig: patches small tank} for $S_{\phi,2}=0.01$, $D_{\phi}=0.001$\ m$^2$\ s$^{-1}$, $\varepsilon_{\phi,0}=0.32$\ rad\ s$^{-1}$. (b): Shown is the intermittent behavior of the CSD (color) over time due to Eq.~\eqref{eq: epsphi2 response}. The green and red dashed vertical lines indicate the time intervals from which the averaged CSDs and ${\cal N}$ are obtained. Color coding indicates the average of $\rho_c(K)$ from $16$ model runs. Time segments concern onset of clustering ($0<t<10$\ s), maintenance of clusters ($30<t<200$\ s) and dissipation of clusters in a saturated tank ($t>200$\ s ). (c). Shown are the CSDs $\rho_c(K)$ for averaging intervals $0<t<10$\ s in black (green dashed lines in (b)) and $30<t<200$\ s in red (red dashed lines in (b)). Additionally $\rho_c(K)$ in an interval $800<t<900$\ s (blue) almost overlaps with the CSD in the beginning of the simulation (black). The bold green line indicates the log-normal distribution having standard deviation $0.45$ and mean $0.3$. Dashed and dashed-dotted vertical lines indicate ${\cal N}$ prior/after and during cluster formation respectively (see Fig.~\ref{fig:MSQ dist tracer} panel (d)).}
    \label{fig: exampe CSD}
\end{figure}

The local mean-square distance, ${\cal R}$, measures the average distance of nearest neighbors, $r$. It indicates when jellyfish swim closer to each other \cite{fily2012athermal,thakur2012collective}. The hexagonal order parameter, Hex, measures the local six-fold geometric bond-order in the swarm. It varies between zero (disorder) and unity (perfect hexatic order in an infinite system) \cite{rex2007lane, gasser2010melting}. Note that different notions of locality have been reported in the literature. Many studies average over $m=6$ nearest neighbors \cite{steinhardt1983bond, bialke2012crystallization}, use the number of direct neighbors in a Voronoi tessellation \cite{gasser2014characterization, zydek2021description, paoluzzi2022motility} or introduce a short distance inside which the averaging takes place \cite{romano2011two, redner2013reentrant, navarro2015clustering}. These studies deal with dense suspensions in which agent-agent interactions dominate and lead to a pronounced ordering effect. As a consequence all of those definitions are equivalent. On the contrary, our simulations concern low-density swarms where the volume exclusion in Eq.~\eqref{eq: positional dynamics} can not establish strong local order. To still capture the effect of occasional volume exclusion, we average over $m>6$ nearest agents. Moreover, drone imaging of jellyfish will always capture the 2D projection of the 3D positions of jellyfish in the upper water layer \cite{rex2007lane, hamel2021using}. In 3D, the optimal number of nearest agents will increase beyond $m=6$ because the surface area (for a unit sphere) of $2\pi \approx 6$ (2D) increases to $4\pi \approx 12$. Thus, $m=24$ is a choice closer to the reality of a future observational tasks. \textcolor{black}{We have additionally checked results based on nearest-neighbor locality by the alternative approach of Voronoi tessellation. In the latter case, the local average runs over different $m$ nearest neighbors (see Fig.~\ref{fig:MSQdist tracer eps dependence }, \ref{fig:MSQ dist tracer}).}

The tracer ratio is an indicator for the search success of the swarm. It measures the excess of the ambient tracer concentration in the swarm, $\phi_J=\langle \phi \rangle_N$ in comparison to the averaged tracer concentration in the tank, $\phi_{\text{tank}}=\langle \phi \rangle_{\text{tank}}$ which would be experienced if jellyfish were purely passive tracers. The latter is obtained by an average over all the grid cells in the fluid domain. The average cluster size $\cal N$ is given by the first moment of the cluster-size distribution (CSD), $\rho_c(K)$. The CSD is calculated according to a geometric interaction radius $R_c$: If $K$ ABP-jellyfish form a geometric graph in which each link is shorter than $R_c$, they belong to the same cluster of size $K$ (see Fig.~\ref{fig: exampe CSD} (a)) \cite{khokonov2021cluster,meakin1985dynamic, theurkauff2012dynamic}. We average the CSD over $M$ model runs and over different time intervals (see Fig.~\ref{fig: exampe CSD} (b,c)). We denote the respective run-ensemble-time averages of all quantities in Eq.~\eqref{eq: statistical quantities} by an overbar.  
    
\subsection{Parameter Settings}\label{sec: params}

The dynamics of jellyfish is specified by $20$ parameters which are listed in Tab.~\ref{tab: params inputs2}. We use here an interaction radius of $R_i=0.15$\ m, in contrast to \cite{gengel2023physics} where $R_i=0.1$\ m. Our intention is to allow for a larger sphere of influence while keeping in mind that the bell diameter is given by $R_{\text{min}} = 0.1$\ m. 

\begin{table}
    \centering
    \begin{tabular}{|ll||l||l|}
     \hline
        Parameter, [unit] & Value & Process & Parameter, [unit] \\
        \hline
        $R_i$, [m] & $0.15$ &  &  \\
        \hline
        $V_0$, [m\ s$^{-1}$] & $0.15$ & Active swimming &  \\
        $J_v$, [-] & $1$ &  &  \\
        $V_1$, [m\ s$^{-1}$] & $0.5$ &  &  \\
        $V_2$, [m\ s$^{-1}$] & $0.6$ &  &  \\
        $S_{\mathbf{U},1}$, [s\ m$^{-1}$] & $400$ &  &  \\
        $S_{\mathbf{U},2}$, [m\ s$^{-1}$] & $0.03$ &  &  \\
        $\varepsilon_{\mathbf{U},0}$, [rad\ s$^{-1}$] & $0.16$ &  &  \\
        \hline
        $S_{{\cal C},1}$, [s\ rad$^{-1}$] & $400$ & Avoidance of turbulence &  \\
        $S_{{\cal C},2}$, [rad\ s$^{-1}$] & $0.005$ &  &  \\
        $\varepsilon_{{\cal C},0}$, [rad\ s$^{-1}$] & $0.08$ & & \\
        \hline
        $\Omega$, [rad\ s$^{-1}$] & $7.54$ & Bell pulsation & \\
        \hline
        $\lambda_{\theta}$, [s$^{-1}$] & $-5$ & Orientation dynamics & $\varepsilon_{\phi,0}$, [rad $s^{-1}$] \\
        $D_{\theta}$, [rad$^{2}$\ s${-3}$] & $0.1$ &  & $S_{\phi,2}$, [$\phi_0$] \\
        $S_{\phi,1}$, [$\phi_0^{-1}$] & $400$ & & \\
        \hline
        $C_{\phi}$, [-] & $0.0083$ (grid 1) & Signaling diffusion & $D_{\phi}$, [m$^2$\ s$^{-1}$] \\
          & $0.00031$ (grid 2) & & $\Gamma$, [s$^{-1}$] \\
        $R_s$, [m] & $1$ (grid 1) & &  \\
          & $2$ (grid 2) & & \\  
        \hline
    \end{tabular}
    \caption{Listing of model parameters. The central column indicates the process in the model. The left column list parameters that are fixed in this model. The right columns list free parameters investigated in this study.}
    \label{tab: params inputs2}
\end{table}

\begin{figure}
    \centering
    \includegraphics[width=\columnwidth]{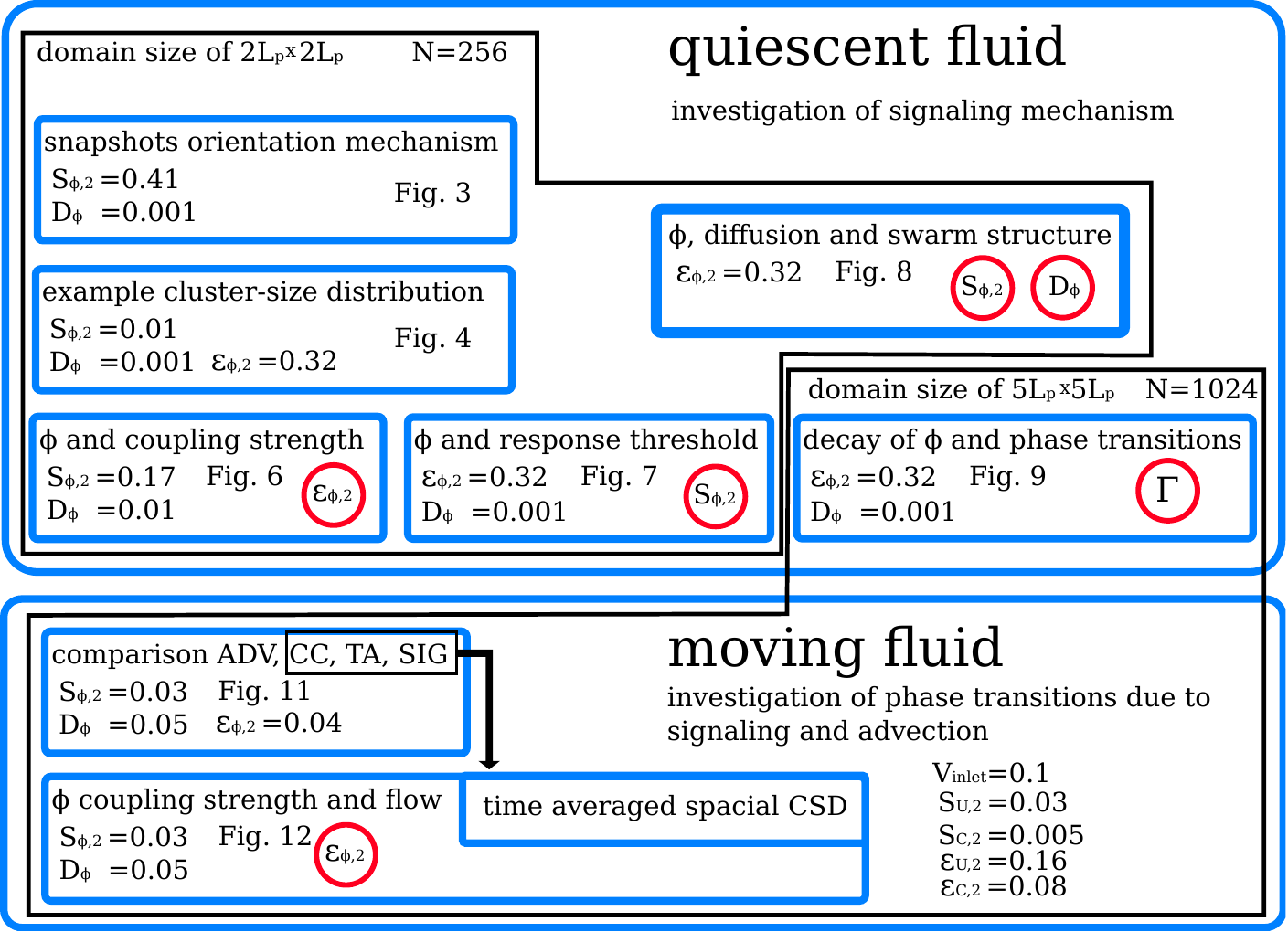}
    \caption{\textcolor{black}{Overview of figures and numerical experiments. Parameter scans are indicated by red circles, fixed parameters are listed. The paper considers a quiescent and a moving flow. The quiescent fluid is used for the numerical study of the signaling mechanism. Investigations of phase transitions are carried out for both flow regimes and on the larger grid with $5L_p \times 5L_p$ to reduce effects of walls. Abbreviations are: advection of active particles with no directional preference (ADV), counter-current swimming (CC), turbulence avoidance (TA), signaling (SIG).}}
    \label{fig: parameters}
\end{figure}

The mixing factor $C_{\phi}$ follows from the assumptions of cubic grid cells and a spherical shape for the jellyfish bell. Then, $\text{Volume}_{\text{bell}} \sim R_{\text{min}}^3$. According to in-situ measurements of the wake volume \cite{dabiri2005flow}, where $3\text{Volume}_{\text{bell}}=1213\cdot 10^{-6}$\ m$^3$ (for $R_{\text{min}}=0.05$\ m) we extrapolate a mixed bell volume (for $R_{\text{min}}=0.1$\ m) $\text{Volume}_{\text{bell}}=1213\cdot 10^{-6}8/3$\ m$^3 = 0.0032$\ m$^3$.
This results in mixing factors for grid 1: $\text{Volume}_{\text{cell}} = (30/41)^3$\ m$^3 \Rightarrow C_{\phi}=0.0083$ and for grid 2: $\text{Volume}_{\text{cell}} = (90/41)^3$\ m$^3 \Rightarrow C_{\phi}=0.00031$. 

\section{Results}\label{Sec: results}

We first \textcolor{black}{discuss} the situation of a quiescent fluid ($\mathbf{U}=0$) without a tracer sink ($\Gamma=0$). The dynamics Eqs.~(\ref{eq: phase unperturbed}-\ref{eg: sink signaling}) in this case still features the free parameters $\varepsilon_{\phi,0}$, $S_{\phi,2}$ and $D_{\phi}$. We first consider a smaller tank with $900$\ m$^2$ area (grid 1). \textcolor{black}{The edge length of this tank is roughly similar to $2L_p$ for the given swimming parameters of Tab.~\ref{tab: params inputs2}. Since also the walls are impermeable for $\phi$, this setting addresses the signaling response of a small part of a larger jellyfish swarm for which the tracer remains on average constant at the boundary of the averaging volume. The ensemble consists} of $N=256$ jellyfish which start in a square of $28 \times 28$\ m$^2$ (area density of $N\pi R_i^2/28^2 =0.023$). We average results here over $16$ model runs. 

\begin{figure}
    \centering
    \includegraphics[width=\columnwidth]{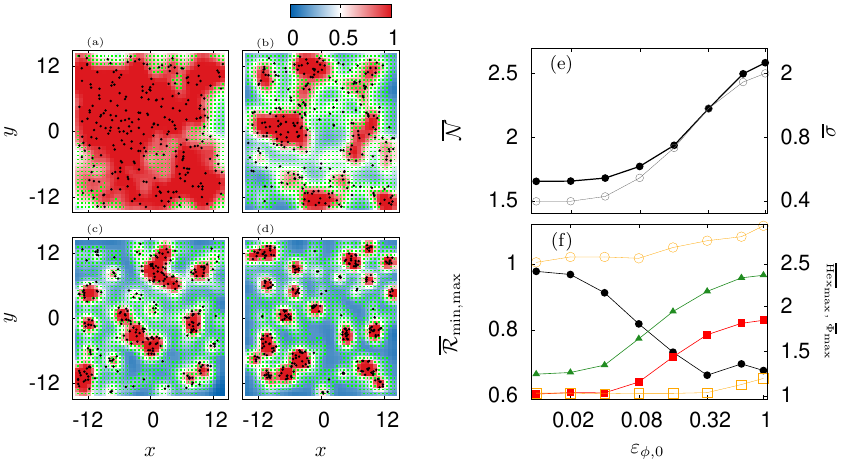}
    \caption{Depicted is the dependence of jellyfish clustering on the orientation strength $\varepsilon_{\phi,0}$. Background colors indicate $\phi$. Green squares label the recruitment domain. (a): $\varepsilon_{\phi,0} = 0.01$\ rad\ s$^{-1}$. (b): $\varepsilon_{\phi,0} = 0.04$\ rad\ s$^{-1}$, (c): $\varepsilon_{\phi,0} = 0.16$\ rad\ s$^{-1}$, (d): $\varepsilon_{\phi,0} = 0.64$\ rad\ s$^{-1}$. (e): Shown is the average cluster size $\overline{\cal N}$ (black) with its standard deviation $\overline{\sigma}$ (grey) for a geometric interaction radius $R_c=0.5$.\ m. (f): Shown is the dependence of the minimal local agent-agent distances ${\cal R}_{\text{min}}$ (black dots, left axis) and the maximal hexagonal order of the ABP-jellyfish (red squares, right axis) and the maximum ratio of the signaling tracer concentration $\Phi$ (green triangles, right axis). Other model parameters are $S_{\phi,2} = 0.17$, $D_{\phi}=0.01$\ m$^2$\ s$^{-1}$. \textcolor{black}{Shown in orange are the maximal agent-agent distance $\overline{\cal R}_{\text{max}}$ (circles, left axis) and $\overline{\text{Hex}}_{\text{max}}$ (squares, right axis).}}
    \label{fig:MSQdist tracer eps dependence }
\end{figure}

\subsection{Coupling Strength}\label{sec: coupling strength}

The coupling strength $\varepsilon_{\phi,0}$ determines how fast ABP-jellyfish turn towards the $\phi$-gradient. In Fig.~\ref{fig:MSQdist tracer eps dependence } exemplary snapshots are depicted. For weak coupling \textcolor{black}{$\varepsilon_{\phi,0}$}, the ABP-jellyfish remain effectively uniformly distributed throughout the simulation (panel (a)) while stronger angular couplings evoke amplification of small density fluctuations and formation of successively more and more pronounced and disconnected patches (panels (b-d)). \textcolor{black}{The reason for this is that the reorientation on time scale $\varepsilon_{\phi,0}^{-1}$ competes with the diffusion of $\phi$. The activation threshold $S_{\phi,2}$ determines how long the reorientation is active, resulting in an effective response. \textcolor{black}{When this response is pronounced enough, the directed swimming results in patches.} Accordingly, panel (a) Fig.~\ref{fig:MSQdist tracer eps dependence } represents a situation where diffusion dominated while in panels (b-d) recruitment dominated.} 

This parameter response is reflected by an increase of the average cluster size, ${\cal N}$ (Fig.~\ref{fig:MSQdist tracer eps dependence } panel (e)), a decrease in the minimal average agent-agent distance, ${\cal R}_{\text{min}}$, and an increase in the maximum values of hexagonal order, Hex$_{\text{max}}$, and tracer ratio, $\Phi_{\text{max}}$ (Fig.~\ref{fig:MSQdist tracer eps dependence } panel (f)). \textcolor{black}{Since there are no $\phi$-sinks present,} we observe that the ABP-jellyfish groups are dissolved after formation because $\phi$ gets enriched and diffused. Accordingly, we use the extreme values of ${\cal R}$, Hex and $\Phi$. For the same reason, the average cluster size was obtained during the clustering period ranging from $30$\ s to $200$\ s (see Fig.~\ref{fig: exampe CSD}). \textcolor{black}{Using Voronoi tessellation, we found that the sensitivity to observed structural changes is significantly reduced (orange lines in panel (f) Fig.~\ref{fig:MSQdist tracer eps dependence }). In particular, the average agent-agent distance reaches a maximum during the clustering period. This is expected because Voronoi cells are convex by definition and are particularly large for agents at the edges of patches such that far away agents are geometrically connected. In contrast by using the $m$ or $n$ nearest neighbors always the closest agents are selected which leads eventually to disconnection of converged clusters in analysis (see App.~C).}

\subsection{Tracer and Clustering}\label{sec: tracer and clustering}

\begin{figure}[!htbp!]
    \centering
    \includegraphics[width=\columnwidth]{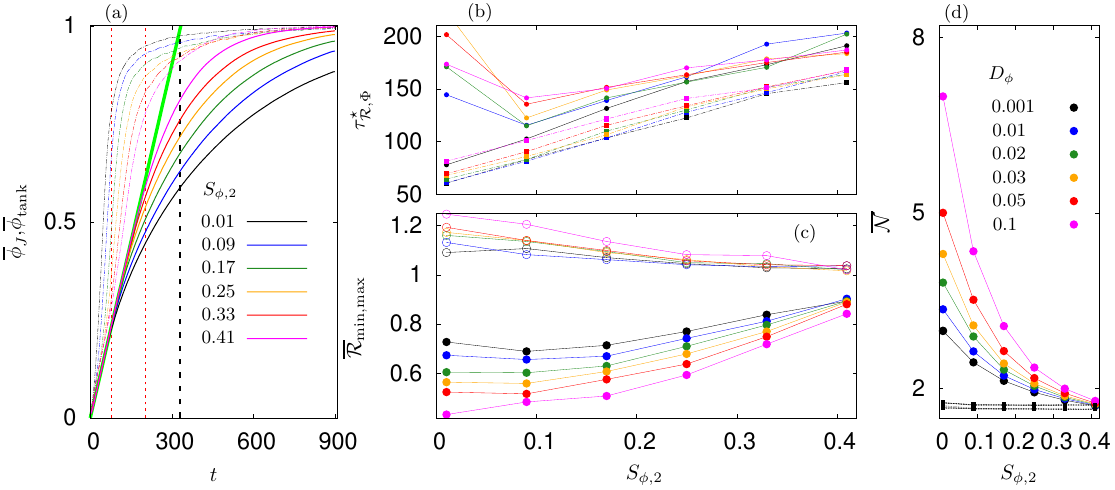}
    \caption{(a): Depicted are the averaged transients for the averaged concentration of agents, $\phi_J$ (dashed lines) and in the tank $\phi_{\text{tank}}$ for different sensitivities $S_{\phi,2}$ and diffusivities $D_{\phi}=0.001$\ m$^2$\ s$^{-1}$, $\varepsilon_{\phi,0}=0.32$\ rad\ s$^{-1}$. The green line indicates the slope ${\cal J}=0.003$\ s$^{-1}$. the vertical black dashed line indicates $\tau_{\text{sat}}$ and the red lines indicate the range of all the clustering times $\tau^{\star}_{\cal R}$. (b): Depicted are the clustering times $\tau^{\star}_{\cal R}$ (bold lines, dots) and the times, $\tau^{\star}_{\Phi}$, when the tracer ratios $\Phi$ reach their maximum (dashed lines, squares). (c): Shown are the dependence of the minimal and maximal local mean square distances, ${\cal R}_{\text{min}}$ and ${\cal R}_{\text{max}}$ on $S_{\phi,2}$ and $D_{\phi}$. (d): Shown is the averaged cluster size $\overline{\cal N}$ during the clustering period, depending on $S_{\phi,2}$ and $D_{\phi}$ for $R_c=0.5$\ m.}
    \label{fig:MSQ dist tracer}
\end{figure}

The tracer dynamics is of further interest. In fact, since we have not included a tracer sink for now ($\Gamma=0$), $\phi$ increases linearly with a rate 
\begin{equation}
{\cal J} =  C_{\phi} \frac{N}{A_{\text{tank}}} \frac{\Omega}{2\pi}\; ,\label{eq: J definition}
\end{equation}
when the ABP-jellyfish are uniformly distributed \textcolor{black}{and when the concentration $\phi$ is still small} (see Fig.~\ref{fig:MSQdist tracer eps dependence } panel (a)). Exemplary transients of $\phi_J$ and $\phi_{\text{tank}}$ are shown in Fig.~\ref{fig:MSQ dist tracer} panel (a). In this simple setting, $\phi_J$ and $\phi_{\text{tank}}$ rise almost linearly with a rate ${\cal J}=0.003$\ s$^{-1}$ prior to patch formation and $\varepsilon_{\phi}({\cal J}t)$ is only a function of time. After structural collapse at approximately $100$\ s the swarm concentration, $\phi_J$, increases more rapidly because the concentration has reached the recruitment threshold for $\varepsilon_{\phi}$ in Eq.~\eqref{eq: epsphi2 response}. Accordingly, jellyfish converge towards local fluctuations of $\rho$ indicated by $\phi$ and the tank concentration, $\phi_{\text{tank}}$, crosses over to a more modest growth rate governed by diffusive transport of $\phi$ to regions between patches, \textcolor{black}{a reduced tracer release inside patches} and release of tracer by solitary jellyfish. After careful checking of our results, we found that most of the density collapse takes place before the linear saturation time $\tau_{\text{sat}} = {\cal J}^{-1}$ as can be seen by a comparison of the range of clustering times $\tau^{\star}_{\cal R}$ (red dashed vertical lines in Fig.~\ref{fig:MSQ dist tracer} panel (a) and solid\textcolor{black}{-dotted} lines in panel (b)). \textcolor{black}{Thus, we consider our saturated tracer release (see App.~B) suitable at least in the recruitment phase}. 

\begin{figure}[!htbp!]
    \centering
    \includegraphics[width=\columnwidth]{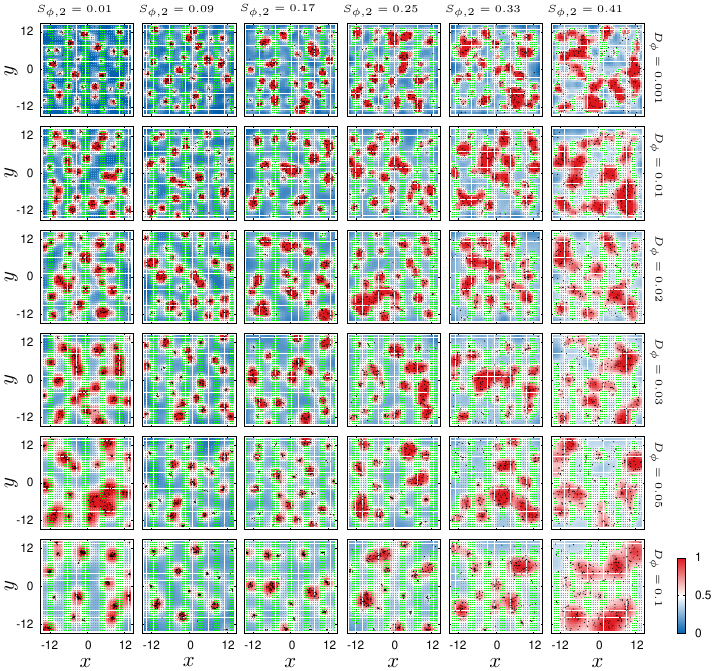}
    \caption{Depicted are snapshots of the patch structures at clustering times $\tau^{\star}$ corresponding to points in Fig.~\ref{fig:MSQ dist tracer} panel (b,c). The angular coupling strength is $\varepsilon_{\phi,0}=0.32$\ rad\ s $^{-1}$. Color maps indicate $\phi$. Black dots indicate the ABP-jellyfish. Regions where the self-induced stimulus affects the angular dynamics are indicated by a green raster.}
    \label{fig: patches small tank}
\end{figure}

We further tested the dependence of the \textcolor{black}{recruitment process on $D_{\phi}$ and $S_{\phi, 2}$. Again, diffusion is responsible for structural smoothing. We fixed $\varepsilon_{\phi,0}$ such that now only $S_{\phi,2}$ determines how decisive the reorientation is.} We see in Fig.~\ref{fig:MSQ dist tracer} panel (c) that ${\cal R}_{\text{min}}$ is inversely proportional to $D_{\phi}$. The reason for this is that under fast structural smoothing small clusters or solitary jellyfish are almost always absorbed into larger clusters. In contrast, weak diffusion hampers the growth of patches such that \textcolor{black}{reorientation can dominate and} solitary medusae or small clusters are more likely. In this case, averages over the $n$ nearest jellyfish become non-local and take into account members of other clusters (compare snapshots Fig.~\ref{fig: patches small tank}). ${\cal R}_{\text{min}}$ mostly scales $\sim S_{\phi,2}$ at low sensitivities ($S_{\phi,2} \rightarrow 0.5$). Interestingly, at very high sensitivities ($S_{\phi,2} \rightarrow 0$) ${\cal R}_{\text{min}}$ is quasi independent of $S_{\phi,2}$, indicating that cluster formation is driven by fluctuations of jellyfish density $\rho$ rather than behavioral sensitivity. Our simulations show that the most clustered state corresponding to ${\cal R}_{\text{min}} = {\cal R}(\tau^{\star}_{\cal R})$ is approached at different times $\tau^{\star}_{\cal R}$. This happens after the maximal tracer ratio $\Phi_{\text{max}}=\Phi(\tau^{\star}_{\Phi})$ was reached at $\tau^{\star}_{\Phi}$ (see Fig.~\ref{fig:MSQ dist tracer} panel (b) where $\tau^{\star}_{\cal R}>\tau^{\star}_{\Phi}$). We attribute the spread of $\tau^{\star}_{\cal R}$ to the process of cluster merging when the clusters are initially quite small.
\textcolor{black}{In Fig.~\ref{fig:MSQ dist tracer} panel (c) we also show the maximal agent-agent distance ${\cal R}_{\text{max}}$ based on Voronoi tessellation. We again see a much weaker response compared to the nearest-neighbors approach.}

Finally, the average cluster size, ${\cal N}$, indicates that the largest clusters form for small $S_{\phi,2}$ (see Fig.~\ref{fig:MSQ dist tracer} panel (d)) and that clusters become bigger when tracer spreads out more rapidly according to larger $D_{\phi}$. When $S_{\phi,2} \rightarrow 0.5$, ${\cal N}$ approaches the cluster sizes found in the quasi-random ensembles at simulation start and end (black square-dashed lines in panel (d)). $\overline{\cal N}$ is obtained from time averages of the CSD, $\rho_c(K)$, in intervals $0$\ s $<t<10$\ s (simulation start), $30$\ s$<t<200$\ s (clustering period), $800$\ s $<t<900$\ s (simulation end). 

\subsection{Effect of a Tracer Sink}\label{sec: tracer sink}

Now, we investigate the effect of a tracer sink ($\Gamma>0$) on patch formation, based on the full diffusion dynamics Eq.~\eqref{eq: diffusion of tracers}. For a uniform tracer field and uniformly distributed jellyfish, spacial averaging yields the tracer concentration 
\begin{equation}
    \phi_{\text{tank}}(t) = \frac{{\cal J}}{\Gamma}\left(1-\exp(-\Gamma t) \right)\; . \label{eq: phi tank gamma}
\end{equation}
It is immediately clear that the saturation value $\phi^{\star}={\cal J}/\Gamma$ potentially lies below the threshold value $S_{\phi,2}$ effectively preventing the onset of patch formation due to the simplified response of Eq.~\eqref{eq: epsphi2 response}. Moreover, the characteristic time scale for tracer dynamics is now $\tau'_{\text{sat}} =\Gamma^{-1}$. We divert our investigations now to a larger tank of $90\times90$\ m$^2$ with $N=1024$ jellyfish initialized in a square of $72\times72$\ m$^2$. 

Due to an expansion of the swarm in the beginning of the simulation, we use the area of $81 \times 81$\ m$^2$ corresponding to the average edge length of $81$\ m. These settings result in $N \pi R_i^2/81^2 = 0.011$, ${\cal J} = 0.000058$\ s$^{-1}$, $\tau_{\text{sat}} \approx 5$\ h. \textcolor{black}{The given tank size corresponds to roughly $5L_p$ such that effects of the boundary on jellyfish swimming are reduced. We accepted a changed in the grid resolution here, mainly due to computational constraints. Since the coupling radius $R_s=2$\ m, an initially stronger tracer mixing is implied. These simulations cover also the much longer time span of two hours which corresponds to at least $4$ characteristic decay times of $\phi$, given by $\Gamma^{-1}$}

We consider the fixed parameters $D_{\phi} = 0.001$\ m$^2$\ s$^{-1}$, $S_{\phi,2} = 0.03$ \textcolor{black}{and $\varepsilon_{\phi,0}=0.32$\ rad\ s$^{-1}$. This resembles the situation of slow lateral diffusion ($L_p^2/D_{\phi} \sim 90$\ h) and much faster time scales of jellyfish swimming and tracer decay. Accordingly, the formation of blooms is induced by behavioral responses of jellyfish and their movement rather than a reaction-diffusion dynamics of the tracer itself.} 

\begin{figure}
    \centering
    \includegraphics[width=\columnwidth]{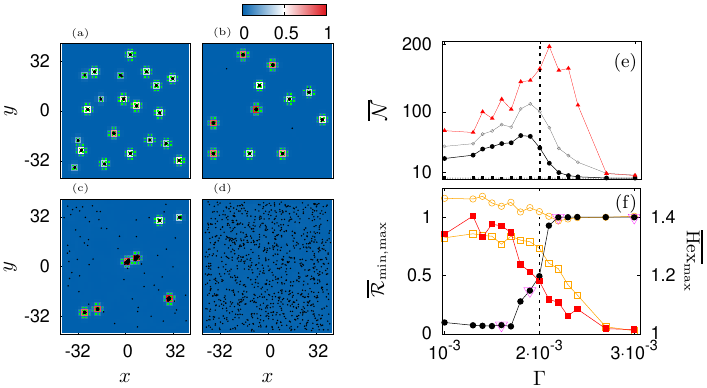}
    \caption{Depicted are snapshots of the swarms at $t=7200$\ s, similar to Fig.~\ref{fig:MSQdist tracer eps dependence }. (a): $\Gamma=0.0016$\ s$^{-1}$, (b): $\Gamma=0.0019$\ s$^{-1}$,  (c): $\Gamma=0.0022$\ s$^{-1}$,  (d): $\Gamma=0.003$\ s$^{-1}$ (pink triangles in panel (f)). Other parameters of the simulation are $\varepsilon_{\phi,0} = 0.32$\ rad\ s$^{-1}$, $D_{\phi} = 0.001$\ m$^2$\ s$^{-1}$, $S_{\phi,2}=0.03$. Shown in panel (e) is the average cluster size $\overline{\cal N}$ (black dots) with its upper standard deviation (grey circles) and the average cluster size in the $95\%$ percentile of $\rho_c(K)$ (red triangles) $R_c=1$\ m. Panel (f) presents the maximal hexagonal order parameter, Hex$_{\text{max}}$ (red squares, right axis), and the minimal local mean square distance, ${\cal R}_{\text{min}}$ (black dots, left axis), \textcolor{black}{both obtained from Euclidean nearest neighbor analysis. Instead, orange curves indicate Hex (squares) and ${\cal R}_{\text{max}}$ (circles) based on Voronoi tessellation}. The dashed vertical line indicates $\Gamma_{\text{crit}}$.}
    \label{fig: msq dist gamma}
\end{figure}

The critical sink strength is given by $\Gamma_{\text{crit}} = {\cal J}/S_{\phi,2} = 0.002$\ s$^{-1}$ (vertical dashed line in Fig.~\ref{fig: msq dist gamma}). We average our results over $4$ runs.

An increase of sink strength, $\Gamma$, causes successively less patches to form, indicating a clustering transition (see Fig.~\ref{fig: msq dist gamma} panels (a-d)). Largest clusters emerge near $\Gamma_{\text{crit}}$ (see Fig.~\ref{fig: msq dist gamma} panels (e)). However, even at sink strength as high as $\Gamma=0.0027$\ s$^{-1} > \Gamma_{\text{crit}}$ clusters can be observed within our simulated time. Moreover, fluctuations of cluster size are largest near the clustering transition as well. This becomes particularly clear if we observe the largest clusters, indicated by the $95\%$ percentile of $\rho_c$ (red triangles in Fig.~\ref{fig: msq dist gamma} panel (e)). A comparison of panels (e) and (f) shows that ${\cal R}$, Hex and ${\cal N}$ all indicate the clustering transition at $\Gamma_{\text{crit}}$. In case of ${\cal R}$, panels (c,d) in Fig.~\ref{fig: msq dist gamma} indicate that a few remaining solitary jellyfish dominate the averaging procedure. On the contrary, Hex shows a much more gradual transition. 

\textcolor{black}{The emergence of large clusters near the clustering threshold $\Gamma_{\text{crit}}$ is of further interest. In those simulations, the balance of tracer sink and induced tracer fluctuations is so delicate that the emergence of a single cluster suffices to deprive its surrounding of jellyfish. Consequently, the equilibrium tracer levels ${\cal J}/\Gamma$ in the domain around a cluster drop and remain below $S_{\phi,2}$ while the cluster itself becomes a strong source of $\phi$ that eventually incorporates the remaining freely swimming jellyfish. This seems to be a reflection of the actual swarm recruitment process in the calm sea or in bays and estuaries which indeed has a strong horizontal swimming component while the swarming inside the cluster can be more complex \cite{ilamner1981long, albert2011s}. Our analysis suggests that larger clusters are quite naturally favored over smaller ones. For this effect to be visible, the initial density of spread out jellyfish needs to be already high enough for local density fluctuation to reach the clustering threshold. Field observations are needed to check this hypothesis on proper spatial scales and by mimicking different stimuli.}

\subsection{Interplay of environmental stimulus, ignorance and coherence}\label{sec: counter flow}

\begin{figure}
    \centering
    \includegraphics[width=\columnwidth]{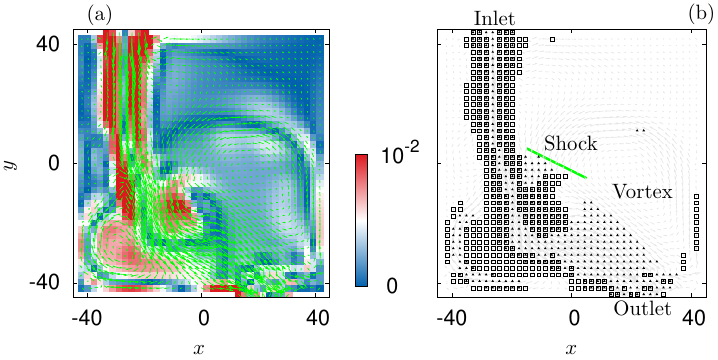}
    \caption{(a): Depicted are typical flow field $\mathbf{U}$ (green arrows) and the absolute vorticity $|{\cal C}|$ (colors) at $4800$\ s. (b): Shown with markers are the decision regions. Triangles indicate regions of counter-current swimming for $S_{\mathbf{U},2}=0.03$\ m\ s$^{-1}$ and boxes indicate regions of turbulence avoidance for $S_{{\cal C},2}=0.005$\ rad\ s$^{-1}$. The green line indicates the position of the turning shock where clusters are formed.}
    \label{fig: example complex flow geometry}
\end{figure}

Up to this point we have discussed structure formation in the presence of a self-induced stimulus $\phi$. However, it will always act in concert with environmental stimuli. Thus, we here elaborate on the emergence of jellyfish patches in a flow ($\mathbf{U} \neq 0$). For this, we consider the commonly encountered experimental setting of an inflow current (see Fig.~\ref{fig: example complex flow geometry}). Confronted with such an environment, jellyfish will not just consider $\phi$ but also other clues which in our model are the flow $\mathbf{U}$, and the absolute vorticity $|{\cal C}|$. We use the latter as a measure of turbulence. We use the larger tank (grid 2) and drive the fluid with an inflow current of $0.1$\ m\ s$^{-1}$. The outflow is situated in the lower right of the tank \textcolor{black}{and acts as the only sink for $\phi$.} This configuration leads to the formation of a strong jet that separates the tank. Left of the current, a narrow quiescent region and a smaller but strong vortex are found. Right of the current, a large counter-clockwise rotating vortex exists. The upper right part of the tank is relatively quiescent (Fig.~\ref{fig: example complex flow geometry} panel (a)). \textcolor{black}{In these simulations, we set $\Gamma=0$.}

In Eq.~\eqref{eq: UC ignorance} we introduce a simple mechanisms for ignorance. This separates the flow domain into several distinct regions according to the parameter levels $S_{\mathbf{U},2}$ and $S_{{\cal C},2}$ (see Fig.~\ref{fig: example complex flow geometry} panel (b)). We fix here $S_{\mathbf{U},2}=0.03$\ m\ s$^{-1}$ such that the ABP-jellyfish counter the current inside the jet and in the branching arms of the vortices. And we let $S_{{\cal C},2}=0.005$\ rad\ s$^{-1}$ such that turbulence avoidance is mostly present in the edges of the jet (triangles and squares in Fig.~\ref{fig: example complex flow geometry} panel (b)). 

\begin{figure}
    \centering
    \includegraphics[width=\columnwidth]{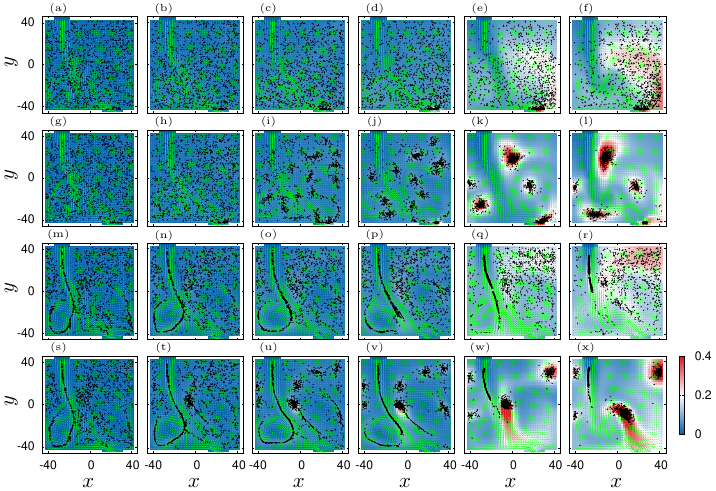}
    \caption{Depicted are swarming behaviors of ABP-jellyfish. Green arrows indicate the flow field. Colors indicate $\phi$. Panels (a-f) show the swarming behavior of the ABP-jellyfish when $\varepsilon_{\phi,0}=\varepsilon_{\mathbf{U},0}=\varepsilon_{{\cal C},0}=0$ (no directional preference, no counter-current swimming). (g-l) shows the swarming transient for $\varepsilon_{\phi,0}=0.04$\ rad\ s$^{-1}$ (signaling attraction, no counter-current swimming). Panels (m-r) show the swarming dynamics for counter-currents swimming and turbulence avoidance. Panels (s-x) show swarm formation for signaling, counter-current swimming and turbulence avoidance. Snapshots are taken at (a,g,m,s): $300$\ s, (b,h,n,t): $600$\ s, (c,i,o,u): $900$\ s, (d,j,p,v): $1200$\ s, (e,k,q,w): $3600$\ s, (f,l,r,x): $4800$\ s. \textcolor{black}{Other model parameters are $D_{\phi} = 0.05$\ m$^2$\ s$^{-1}$, $S_{\phi,2} = 0.03$, $S_{\mathbf{U},2}=0.03$\ m\ s$^{-1}$ and $S_{{\cal C},2}=0.005$\ rad\ s$^{-1}$.}}
    \label{fig:examples complex flow densities a}
\end{figure}

\textcolor{black}{The first setting we discuss here is a swarm of ABP-jellyfish with no directional preference and no adaption of swimming speed (see Fig.~\ref{fig:examples complex flow densities a} panels (a-f)). In this case, the active movement averages out and passive advection results in a net transport to the lower right corner. There, many jellyfish get stuck directly in front of the outlet due to the strong converging current and the counter-acting repulsion force $\mathbf{f}_{\text{ext}}$.}

\textcolor{black}{Next, we allow for a directional preference given by the signaling tracer $\phi$ (see Fig.~\ref{fig:examples complex flow densities a} panels (g-l)). We fix $D=0.05$\ m\ s$^{-2}$, $S_{\phi,2}=0.03$ and $\varepsilon_{\phi,0}=0.04$\ rad\ s$^{-1}$. In this case, we observe that the swarm after a period of random swimming starts to form more and more clusters. These clusters then get transported by the passive flow. The clusters start to act as strong sources of $\phi$ and form separate environments that locally recruit jellyfish. These clusters eventually merge to form larger clusters. The movement of the clusters is then mainly coupled to the passive advection of the tracer $\phi$. }

\begin{figure}
    \centering
    \includegraphics[width=\columnwidth]{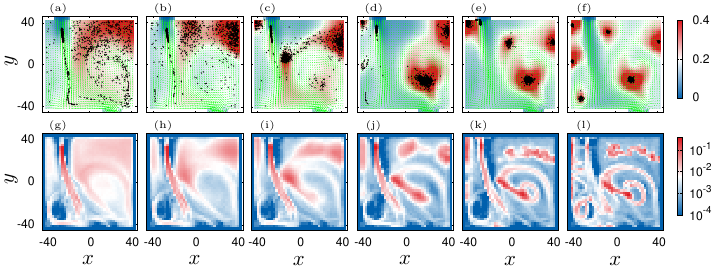}
    \caption{Panels (a-f) shows the swarm in the end of the simulation at $7200$\ s. (a): $\varepsilon_{\phi,0}=0$, (b): $\varepsilon_{\phi,0}=0.01$\ rad\ s$^{-1}$, (c): $\varepsilon_{\phi,0}=0.02$\ rad\ s$^{-1}$, (d): $\varepsilon_{\phi,0}=0.04$\ rad\ s$^{-1}$, (e): $\varepsilon_{\phi,0}=0.08$\ rad\ s$^{-1}$ and (f): $\varepsilon_{\phi,0}=0.16$\ rad\ s$^{-1}$. \textcolor{black}{Other model parameters are $D_{\phi} = 0.05$\ m$^2$\ s$^{-1}$, $S_{\phi,2} = 0.03$, $S_{\mathbf{U},2}=0.03$\ m\ s$^{-1}$ and $S_{{\cal C},2}=0.005$\ rad\ s$^{-1}$.} Panels (g-l) show the corresponding averaged jellyfish densities $\overline{\rho}$.}
    \label{fig:examples complex flow densities b}
\end{figure}

\textcolor{black}{As a next case, we consider counter-current swimming and vorticity avoidance while switching off the orientation due to $\phi$ (Fig.~\ref{fig:examples complex flow densities a} panels (m-r)). In this case, many jellyfish group in front of the inlet because they swim against the current and will avoid the region of strong shear. The transient structure towards the inlet-cluster is a long filament. An other large group instead remains for a long time in the quiescent region of the upper right. This behavior of jellyfish is reflected in the finite-time averaged distribution $\overline{\rho}(\mathbf{x}) = \overline{\langle \rho(\mathbf{x}) \rangle_t}$, depicted in panel (g) of Fig.~\ref{fig:examples complex flow densities b}. There, we clearly see that the jet cluster is the most pronounced feature besides a large density plateau in the upper right of the tank.}

Notably, we observe the tendency of jellyfish to form a loose conglomerate in the center of the tank. The reason for this is a \textcolor{black}{behavioral} turning shock, where the upper arm of the main vortex joins the jet again (green line in Fig.~\ref{fig: example complex flow geometry} panel (b)). There, jellyfish are transported passively from the right into the jet region because they ignore the direction of flow. Once inside the jet, they orient against the current again and encounter more jellyfish coming from the right such that a convergence of medusae is generated. At some point, the emerging cluster falls apart and the jellyfish get transported further downstream. 

\textcolor{black}{Finally, we consider the full orientation dynamics Eq.~\ref{eq: angular coupling} (see Fig.~\ref{fig:examples complex flow densities a} panels (s-x)). In this case, clusters form and are also maintained (compare Fig.~\ref{fig:examples complex flow densities a} panels (g-l) and (s-x)). However, large clusters after the transient are found in different locations: The inlet-cluster is still present and the main clusters has formed at turning-shock while previously clusters form at the outlet, and inside the quiescent vortex centers of the background flow.} The area density of clusters depends on $\varepsilon_{\phi,0}$ (see Fig.~\ref{fig:examples complex flow densities b} panels (a-f)). In fact, $\overline{\rho}$ allows us to identify regions in the flow that are likely to host clusters (see Fig.~\ref{fig:examples complex flow densities b} panels (g-l)). 

\textcolor{black}{We make the following observations:
\begin{itemize}
    \item \textit{Turning shock}: Jellyfish get enriched in the region of the turning shock when counter-current swimming is taken into account. This region acts as a nucleus for later clustering when orientation due to $\phi$ is active. Otherwise clusters eventually fall apart again. Structurally stable patches which have formed in this region get transported downstream with the flow (diagonal structure in the center of panels (j-l) Fig.~\ref{fig:examples complex flow densities b} following the left edge of the main vortex). 
    \item \textit{Vortex center}: The coil structure near the turning shock does not extend or smear out to the right edge of the tank. The reason for this is that $\phi$ gets enriched in the quiescent central region of the main vortex where jellyfish then tend to stay due to the attractive action of $\phi$. \textcolor{black}{This effect is even stronger when only the attraction due to $\phi$ affects the orientation $\theta$.}
    \item \textit{Focus and distraction}: When turbulence avoidance is present, the strong main current causes the ABP-jellyfish to focus in an equilibrium point in front of the inlet where the active swimming of single individuals gets hampered due to non-synchronous pulsation of the bells inside the cluster. The cluster as a whole is then unable to overcome the background flow \textcolor{black}{and remains in front of the inlet}. Stronger coupling to $\phi$, causes the cluster inside the jet to become less prominent because individuals get distracted from their counter-current swimming by orientation towards $\phi$. \textcolor{black}{If jellyfish have no directional preference, they get stuck in front of the outlet due to the outflow current.}
\end{itemize}}

\section{Conclusion}\label{sec: conclusions}

In this paper, we have introduced a theoretical paradigm for jellyfish swarm maintenance based on active Brownian particle simulations. We devised a simplistic mechanism to model jellyfish reaction to walls or coast lines and we introduced three novel parameter responses accounting for ignorance of environmental drivers and self-induced stimuli. We exemplified the interplay of those drivers at the example of a generic signaling tracer, counter current swimming and turbulence avoidance. In doing so we are able to generate swarming behavior well known from observations of jellyfish. 

We have discussed geometric measures for swarm detection of which the cluster size distribution, $\rho_c$, appears to be most suited for further observations. \textcolor{black}{Nearest-neighbor distance and hexagonal order are also suitable to distinguish blooming from free swimming. However, they depend on the definition of locality. We have used the widespread approach of Voronoi tessellation and a more simplistic approach based on Euclidean distance. We consider the latter to be more suitable. Also, because it naturally discards many physically irrelevant long-range connections.}

\textcolor{black}{Our analysis suggests that jellyfish blooms are formed due to a mutual interplay of environment and behavior. This is one of the major differences to other studies on active matter and phase transitions where the background medium is often assumed quiescent! Several scenarios beyond the scope of this paper are conceivable:
\begin{itemize}
    \item The interplay of environmental drivers such as $\mathbf{U}$ or the abundance of prey can cause a local increase of $\rho$ such that the local tracer flux ${\cal J} \sim N \int \rho(\mathbf{x}') dA'(\mathbf{x}) /A(\mathbf{x})$ increases and $\phi$ at position $\mathbf{x}$ surpasses the behavioral sensing threshold $S_{\phi,2}$. The resulting swarm coherence (turning shock in Fig.~\ref{fig: example complex flow geometry}) then acts as a nucleus for swarm formation. This effect should be particularly prominent at the sea surface where in addition to a behavioral shock, the local convergence and divergence of the surface flow is induced by upwelling and downwelling \cite{cressman2004eulerian}.
    \item Jellyfish can also cluster mutually as a consequence of volume exclusion, once $\rho$ has increased sufficiently. In that case, jellyfish start to feel each other and slow down. This leads to an effectively negative swimming pressure, forcing more and more jellyfish into structural collapse \cite{takatori2014swim}. This scenario might be relevant within already formed patches where a stimulus like $\phi$ is maxed out, or when environmental conditions change on time scales much larger than the time scale of active swimming. For example, if $\langle V \rangle_t = 0.067$\ m\ s$^{-1}$ \cite{malul2019levantine}, converging on the scale of a kilometer requires a calm sea for approximately $4$\ h, opting for night-time conditions in which such an effect might be observable. Alternatively this effect could be of interest for closed habitats like Lake Palau \cite{cimino2018jellyfish} and in tank experiments \cite{mackie1981swimming}. 
    \item Jellyfish can change the amount of tracer that is released into the water, subsequently changing the mixing factor $C_{\phi}$ according to their life stage and the season. Above, we also indicated that the mixing factor $C_{\phi}$ is sensitive to the effective fluid volume into which the tracer is released. Consequently, a direct influence of marine topography on blooming could be of relevance \cite{gershwin2014dangerous}. On shortest time scales, the mixing of tracer into the fluid depends on kinematic parameters like $\Omega$, the bell shape and the form of wakes which are not resolved in our simulations.
    \item The interplay of sensing thresholds $S_{\mathbf{U}, {\cal C}, \phi;2}$ influences which $\phi$ concentration triggers the simplified clustering instability described in this study. We have shown that the clustering transition takes place in a quite narrow range of small valued parameters (either $\Gamma$ or $S_{\phi,2}$), in accordance with the observation that jellyfish are extremely sensitive to changes in their environment \cite{albert2011s}. On the one hand, such sensitivity certainly can be attributed to the process of natural selection which should have lead to sharp responses (narrow parameter rage, large value of $p_1$) at very small stimulus amplitudes (see Eq.~\eqref{eq: UC ignorance}) \cite{hamner2009review}. On the other hand, we have seen in Sec.~\ref{sec: coupling strength} that the clustering response at very high sensitivities ($S_{\phi,2} \rightarrow 0$) depends mainly on the diffusivity $D_{\phi}$, emphasizing the importance of externally induced density fluctuations on cluster formation. This suggest a less strict evolutionary corridor for the development of a certain high sensitivity because shock regions like in Fig.~\ref{fig: example complex flow geometry} are a common feature in many flow settings.
\end{itemize}}

Different aspects of model development should be addressed in the future: As Jellyfish are well known to swim in the vertical \cite{kaartvedt2007diel, mackie1981swimming} a three-dimensional swarming dynamics should be devised. The commonly calculated large eddy viscosity could be used as an alternative environmental clue for turbulence \cite{maulik2016dynamic, maulik2017novel}. Moreover, our model has excluded physiological features like positional preferences among medusae \cite{calovi2014swarming}, neurological response mechanisms and the shape and dynamics of wakes. Neuronal responses can be included into the model by introduction of perturbation terms to the phase dynamics Eq.~\eqref{eq: phase unperturbed} \cite{schultheiss2011phase,pikovsky2001synchronization}. Wakes require further numerical and experimental analysis because they might act as a time-delayed means of chemical and kinematic communication on the scale of tanks. Obviously, in ocean simulations, scale separation allows to incorporate those effects into the signaling tracer discussed here. 
\textcolor{black}{Considering the parameters of Tab.~\ref{tab: params inputs2}, we see that jellyfish seem to resemble some kinematic properties of bacteria: They, usually appear in huge numbers, have a limited physiology and feature a rotational persistence number, Pe$= L_p/(2R_{\text{min}}) \sim 85$ \cite{gengel2023physics, saragosti2012modeling,buttinoni2012active,ariel2018collective}.
At the same time they are as big as fish and submerged in flow structures that are much larger in scale. This could render jellyfish an interesting subject of study because they allow to observe otherwise microscopic processes in a dynamically similar macroscopic system. Our results also show that distribution patterns are quite sensitive to parameter. Further model-oriented experiments in tanks or inference from observations in different environmental conditions are needed.}

\section{Acknowledgement}

E.G. thanks Arkady Pikovsky, Alexei Krekhov and Subhajit Kar for helpful discussions. E.G. thanks the Minerva Stiftungsgesellschaft fuer die Forschung mbH and the Max-Planck society for their support. Eyal Heifetz and Zafrir Kuplik are grateful to the Israeli Science Foundation grant 1218/23

\bibliography{manuscript2.bib}

\section*{Appendix A -- Turning at Walls} \label{app A}

Turning at walls is realized by the coupling term $\pi {\cal B}(\mathbf{x}_j)\sin(\theta_j-\delta_{n,j})$, involving the Boolean mappings ${\cal B} \in \{0,1\}$ and $\delta_{n,j} \in \{u \pi/2; u=1, 2, \ldots, 7 \}$, having the form:
\begin{equation}
    \begin{aligned}
        \delta_{n,j} &= \pi \left( B_{j,0}/4 - B_{j,1}/4 + B_{j,2}5/4 - B_{j,3}5/4 + B_{j,0} B_{j,3} 3/2 - B_{j,1} B_{j,2} 3/2 - B_{j,2} B_{j,3} \right) \\
        {\cal B}(\mathbf{x}_j) &= B_{j,0} + B_{j,1} + B_{j,2} + B_{j,3} + B_{j,0}(B_{j,1}(-1+B_{j,2}+B_{j,3})+B_{j,2}(-1+B_{j,3})-B_{j,3}) \\
        & - B_{j,1}(B_{j,2}+B_{j,3}) - B_{j,2}B_{j,3} + B_{j,1}B_{j,2}B_{j,3}\; .
    \end{aligned}
\end{equation}
In these equations the boundary map $B_j \in [0,1]^4$ occurs which assigns to each agent $j$ the status of its four nearest grid points in the swarming domain. When a grid point is a solid cell, the mapping of that point will have a value of $1$. Otherwise, its value is zero.

\section*{Appendix B -- Tracer source term}

The coupling term at an Eulerian position $\mathbf{x}$ in Eq.~\eqref{eg: sink signaling} is given by
\begin{equation}
\begin{aligned}
    \textcolor{black}{\text{Coupling}(\rho,\phi) = H(\Delta )c + H(- \Delta )(1-\phi)} \\
    c(\rho, \mathbf{x}) = \sum_{j=1}^{N(\mathbf{x})} \sum_{t=1}^{N_t} \mathds{K}(\mathbf{x}_j(t),\mathbf{x}) = N \int_{t}^{t+dt} \int_{|\mathbf{x}'-\mathbf{x}|\leq R_s} \mathds{K}(\mathbf{x}',\mathbf{x}) \rho(\mathbf{x}',t') dA'(\mathbf{x}) \delta(\varphi(\mathbf{x}',t'))) dt' \; .
\end{aligned}
\end{equation}
\textcolor{black}{$H(.)$ is the Heaviside function given by 
\begin{equation}
H(a) = \begin{cases} 1, \text{if }\; , a>0 \\ 0, \text{if }\; , a \leq 0 \end{cases} \; .
\end{equation}}
$N(\mathbf{x})N_t$ is the number of \textcolor{black}{bell closures (at $\varphi_j =0$) by} agents inside a coupling radius of $R_s$ around $\mathbf{x}$ \textcolor{black}{during the small time interval $dt$, $\Delta=1-\phi-C_{\phi}c$} and 
\begin{equation}
    \mathds{K}(\mathbf{x}',\mathbf{x}) = \frac{1}{\pi R_s^2} \exp \left(-(\mathbf{x}'-\mathbf{x})^2/R_s^2 \right), \qquad \rho(\mathbf{x},t) = \frac{1}{N}\sum_{j=1}^N \delta(\mathbf{x} - \mathbf{x}_j(t)) \; .
\end{equation}
$\delta(\mathbf{x} - \mathbf{x}_j)$ is Dirac's delta function. Integration then yields
\begin{equation}
\int_{0}^{T} \int_{|\mathbf{x}-\mathbf{x}'|<R_s} c(\rho,\mathbf{x}') dA'(\mathbf{x})dt' = N(\mathbf{x})N_T, \qquad \int \rho(\mathbf{x}') dA_{\text{tank}} = 1\; .
\end{equation}
\textcolor{black}{The combination of $H(.)$ ensures that near $\phi=1$, the injected concentration $C_{\phi} c$ can not increase $\phi$ beyond $1$.}

\section*{Appendix C -- Voronoi tessellation and nearest neighbors}

\begin{figure}[!h!]
\centering
\includegraphics[width=\columnwidth]{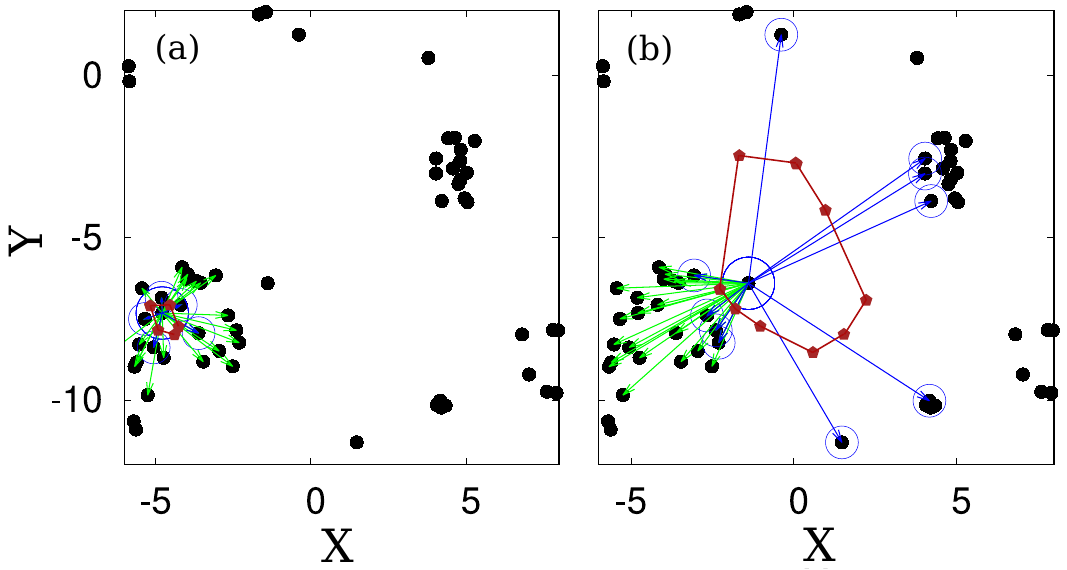}
\caption{Depicted are two instances of Euclidean nearest neighbor selection (green arrows) and Voronoi tessellation (blue arrows). The Voronoi cells are shown in brown. The agent of origin is marked with a larger blue circle. Other agents are shown with black dots. (a): A case where both selection methods yield similar localities. (b): For the case of an edge agent, the Voronoi cell is very large and incorporates spurious connections.}
\label{fig: voronoi nearest neigbors explain}
\end{figure}

\textcolor{black}{Here we elaborate on the problem of neighborhood selection in the statistical analysis. The nearest neighbor approach selects local agents based on their Euclidean distance (green arrows in Fig.~\ref{fig: voronoi nearest neigbors explain}). For the Voronoi tessellation, we used Delaunay triangulation to obtain the set of neighbors (blue arrows in Fig.~\ref{fig: voronoi nearest neigbors explain}). }

\textcolor{black}{Inside dense groups, both approaches yield a similar locality (see panel(a)) for Hex and ${\cal R}$. For solitary agents, both approaches generate connections to far away agents. These connections are somewhat spurious because of the physical limitations of jellyfish. At the edges of clusters, Euclidean distancing selects closest agents inside the cluster while Voronoi tessellation again creates long-range connections. Thus Euclidean distancing is more suitable than Voronoi tessellation (see panel (b)) to calculate ${\cal R}$ while Voronoi tessellation is more suitable for Hex in case of edge agents. Both approaches have to deal with a significant amount of disorder. Our results so far have shown that Euclidean distancing results in better discrimination of clustered and non-clustered states. Our results also suggests to use Euclidean nearest neighbors to calculate Hex since discrepancies of both approaches for edge cases have a minor effect.}

\end{document}